\documentclass[intlimits,twoside,a4paper]{article}
\usepackage[cp1251]{inputenc}
\usepackage{wrapfig}

\usepackage{cmpj3}

\issue{2018}{21}{2}{22801}
\doinumber{10.5488/CMP.21.22801}

\title[Scientometric analysis of Condensed Matter Physics journal]{Scientometric analysis of Condensed Matter Physics journal\thanks{This is an invited article that contains a scientometric analysis of the 25-year publication activities of ``Condensed Matter Physics'' journal.}}
\author{O. Mryglod\refaddr{a1,a2}}
\addresses{\addr{a1}Institute for Condensed Matter Physics of the National
Academy of
Sciences of Ukraine, \\1 Svientsitskii St., 79011 Lviv, Ukraine%
\addr{a2}$\mathbb{L}^4$ Collaboration \& Doctoral College
for the Statistical Physics of Complex Systems,\\
Leipzig-Lorraine-Lviv-Coventry, Europe%
}
\date{Received May 21, 2018, in final form May 30, 2018}

\begin{document}

\maketitle
\begin{abstract}
The paper is dedicated to 25th anniversary of \emph{Condensed Matter Physics} journal (CMP). It contains the results of comprehensive analysis of different journal-related data. CMP co-authorship relationships are studied analysing the collaboration network. Its cumulative statical and dynamical properties as well as the structure are discussed. The international contribution to the journal is assessed using the authors' affiliation data. The network of the countries collaborating within CMP is considered. Another kind of network is used to investigate the topical spectrum: two PACS indices assigned to one paper are connected by link here. The structure of the most significant interdisciplinary connections is analysed. Finally, the download statistics and the corresponding records of the papers' citations are used to discuss the journal's impact.


\pacs 89.75.-k, 02.10.Ox, 02.50.-r, 07.05.Kf

\end{abstract}

\section{Introduction}
25 years ago, in 1993, the first issue of \emph{Condensed Matter Physics} (CMP) journal was printed  \cite{CMP}. Established as three-lingual institutional edition aimed at publishing the results of the research primarily in condensed matter theory representing mainly west Ukrainian authorship, it developed into an authoritative international journal with a powerful expert pool and wide geography. Modern CMP journal not only exhaustively covers the primarily chosen topical directions but also publishes the results in adjacent areas. Today it is recognized by leading scientometric services and covered by information resources and databases. In 2008 the first Journal Impact Factor was defined, but the formation of a scientific edition, visible and accessible to a wide audience and offering the materials of the highest quality, started much earlier. The basis of any good journal --- an independent peer-review procedure assuring in the commonly-accepted way a competitive level of the content --- is provided by experts from all over the world according to a semi-blind procedure. It should be noted that the journal is entirely free of charge for the authors of the submitted papers, while the editorial board besides the technical editing also takes care of the appropriateness of the English writing of the submitted papers by scrupulous proofreading. CMP was the first Ukrainian journal to become available online from the very start. The following steps on the way to promotion, expansion of the readership and increasing the visibility were accomplished. Today, CMP is a member of the Directory of Open Access Journals (DOAJ),  it is regularly reposited into the open e-print archive arXiv.org, all its publications are labelled with DOI numbers, etc. In other words, CMP supports the so-called Diamond Open Access model --- all the materials are prepared, published and made accessible to scientific community by the publisher for free \cite{DiamondOA}. A lot of efforts are made to keep and  strengthen the journal's positions. In particular, a comprehensive database was organized a decade ago to monitor the publishing process using the results of bibliographic data analysis \cite{Mryglod2007}. External data such as citation or download statistics are also used in order to navigate the editorial ship, to correct and improve its strategy. The CMP editorial board is doing a great routine job to publish a good journal and it is an honour to be a part of this team and of the process proper.

This paper is dedicated to 25th anniversary of the journal. It contains the results of scientometric analysis of CMP. Such a research that concentrates on the scale of a single periodical can be considered as a case study. On the other hand, such a descriptive scientometric turns out to be a popular topic.  Quantitative tools can be used to draw a portrait of a scientific discipline or topic \cite{discipline_example,topic_example}, an individual researcher \cite{individual_example}, a journal or a group of scientific journals \cite{journals_example}, countries \cite{country_example}, and so on. In particular, there are several examples of such analysis related to Ukraine: \cite{Mryglod2007,ukraine_example1,ukraine_example2,ukraine_example3,ukraine_example4,jps}.

Here, the main purpose is to draw a general portrait of the journal using different kinds of data. The internal data set consist of bibliographic and editorial records regarding the publications. The Scopus database is used as an external source of information on the citations. The statistics of the visiting journal's website is provided by Google Analytics tools. The quantitative characteristics of CMP authorship are discussed in section~\ref{sec_authorship} along with the analysis of interconnections between the data using a complex network approach. The geography of CMP journal is investigated in section~\ref{sec_geography} --- the co-authorship network at the level of countries is studied therein. A disciplinary landscape of CMP is the subject of section~\ref{sec_PACS}: the information on the mentioning of topical indices in the papers is used. Section~\ref{sec_downloads} considers the statistics of the downloads of CMP papers from its website and makes an attempt to compare them with the statistics of external citations.

\section{Authorship and co-authorship}
\label{sec_authorship}
 Approximately 130 authors publish about 57--58 papers per year in the CMP journal, which makes up 1282 publications in total. This gives us a list of over 1.5 thousand different authors (1666, to be exact) that contributed into CMP during 25 years (1993--2017).
 Approximately 20\% of all the papers are written solely by one author (this number  slightly decreases in time: 25\% for the period 1993--2005 and 20\% for 2006--2017) and 13 papers belong to collaborative teams of more than 6 people. However, the average number of co-authors for CMP is 2~persons, which typically characterizes a theoretical research~\cite{Newman2001,Newman2004}.

\begin{figure}[!t]
\vspace{-7mm}
\centerline{\includegraphics[width=9.0cm]{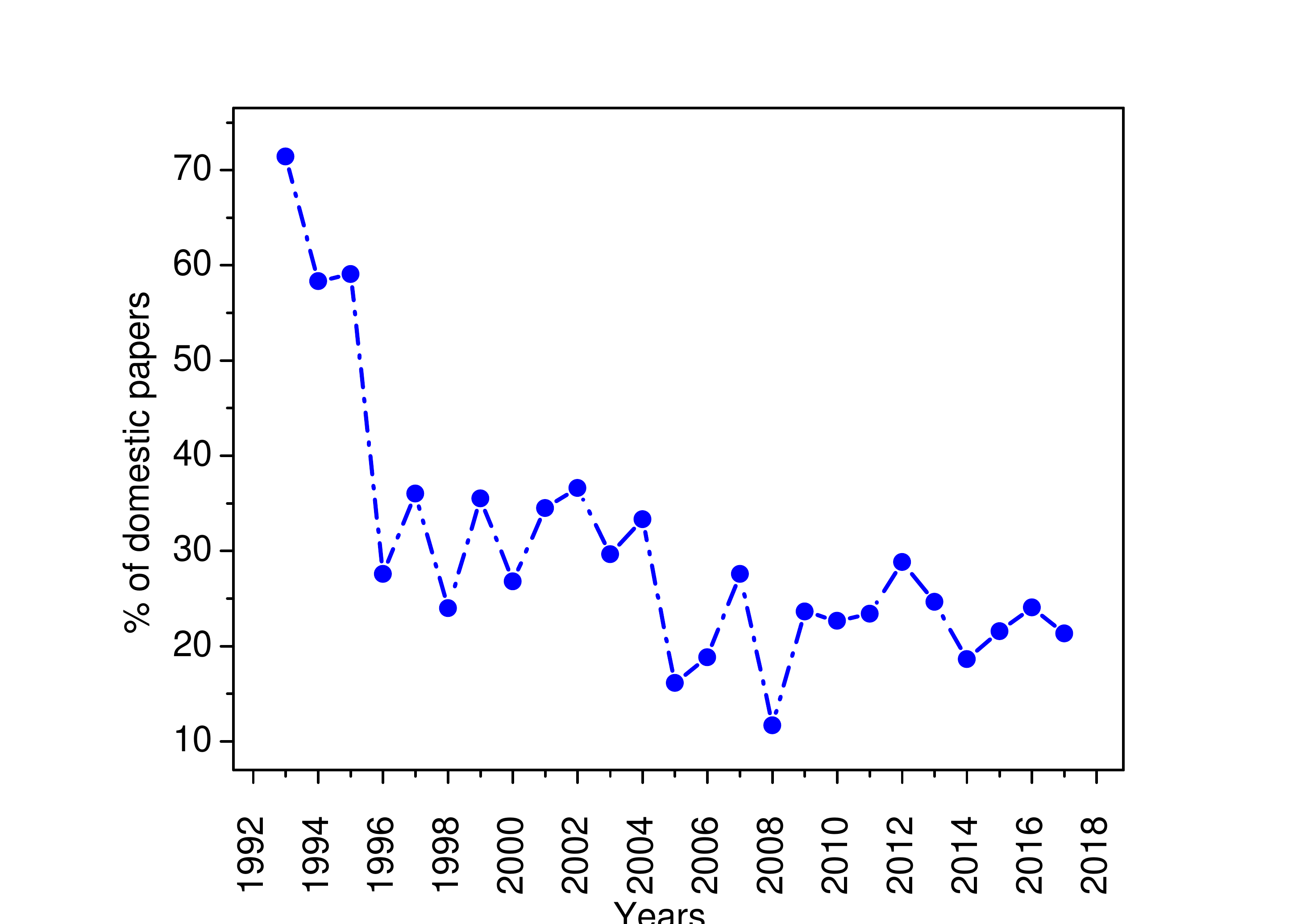}}
\caption{(Colour online) Annual shares of domestic papers (affiliation list contains ICMP) in the CMP journal.}
\label{fig_icmp_papers}
\end{figure}

The frequency distribution of the numbers of CMP publications per one author follows the power law.
On the one hand, 1183 authors (71\%) have contributed into CMP only once and, therefore, they cannot be considered as representative ones. Only approximately 1/6 of the authors with a single paper participate in the largest collaborative group (LCC, see below). On the other hand, there are a few CMP authors that published many papers, see table~\ref{tab_coauth_TOPS}. The highest five positions by the number of CMP publications are occupied by the authors from the journal's domestic institute --- Institute for Condensed Matter Physics (ICMP). However, the situation becomes different if one looks at the rating built for the recent 10 years, see the second column in table~\ref{tab_coauth_TOPS}. The annual shares of the papers where at least one affiliation is ICMP are shown in figure~\ref{fig_icmp_papers}. As one can see, only 1/5 of publications for the last ten years is related to ICMP. A decrease of this share apparently means that more foreign contributions can be potentially published which would increase the journal's international openness~\cite{Jesus2004}.

\begin{table}[!t] 
\caption{TOP5 positions of the ratings of CMP authors by the number of papers (1st and 2nd columns) and by the total number of collaborators (3rd column). The authors from the journals' domestic institute (ICMP) are highlighted in bold. }%
\vspace{2ex}
\begin{center}
 \label{tab_coauth_TOPS}
\begin{tabular}{|c|c|c|c|}
\hline\hline \parbox[t]{0.25\textwidth}{Number of papers \par (entire period of 25 years)}&\parbox[t]{0.25\textwidth}{Number of papers \par (last 10 years) }&
\parbox[t]{0.25\textwidth}{ Number of collaborators \par (entire period of 25 years)}\\\hline\hline
\parbox[t]{0.25\textwidth}{ 43 \textbf{I.V.~Stasyuk}\par
40 \textbf{R.R.~Levitskii}\par
36 \textbf{M.F.~Holovko}\par
31 \textbf{I.M.~Mryglod}\par
31 \textbf{M.V.~Tokarchuk}\par
23 \textbf{Yu.~Holovatch}}
&
\parbox[t]{0.25\textwidth}{
14 \textbf{R.R.~Levitskii}\par
14 O.~Pizio\par
14 S.~Sokolowski\par
13 \textbf{I.V.~Stasyuk}\par
12 D.J.~Henderson\par
12 \textbf{M.F.~Holovko}\par
12 O.M.~Voitsekhivska\par
12 I.R.~Zachek\par
10 \textbf{Ja.M.~Ilnytskyi}\par
10 A.~Patrykiejew\par
10 Ju.O.~Seti\par
10 \textbf{A.S.~Vdovych}\par
9 M.V.~Tkach\par}
&
\parbox[t]{0.25\textwidth}{
27 \textbf{I.V.~Stasyuk}\par
26 \textbf{R.R.~Levitskii}\par
24 \textbf{M.V.~Tokarchuk}\par
23 \textbf{M.F.~Holovko}\par
23 Yu.M.~Vysochanskii\par
20 \textbf{Yu.~Holovatch}\par
20 \textbf{A.D.~Trokhymchuk}}
\\\hline\hline
\end{tabular}
\end{center}
\end{table}

\begin{figure}[!t]
\centering
\includegraphics[width=0.48\textwidth]{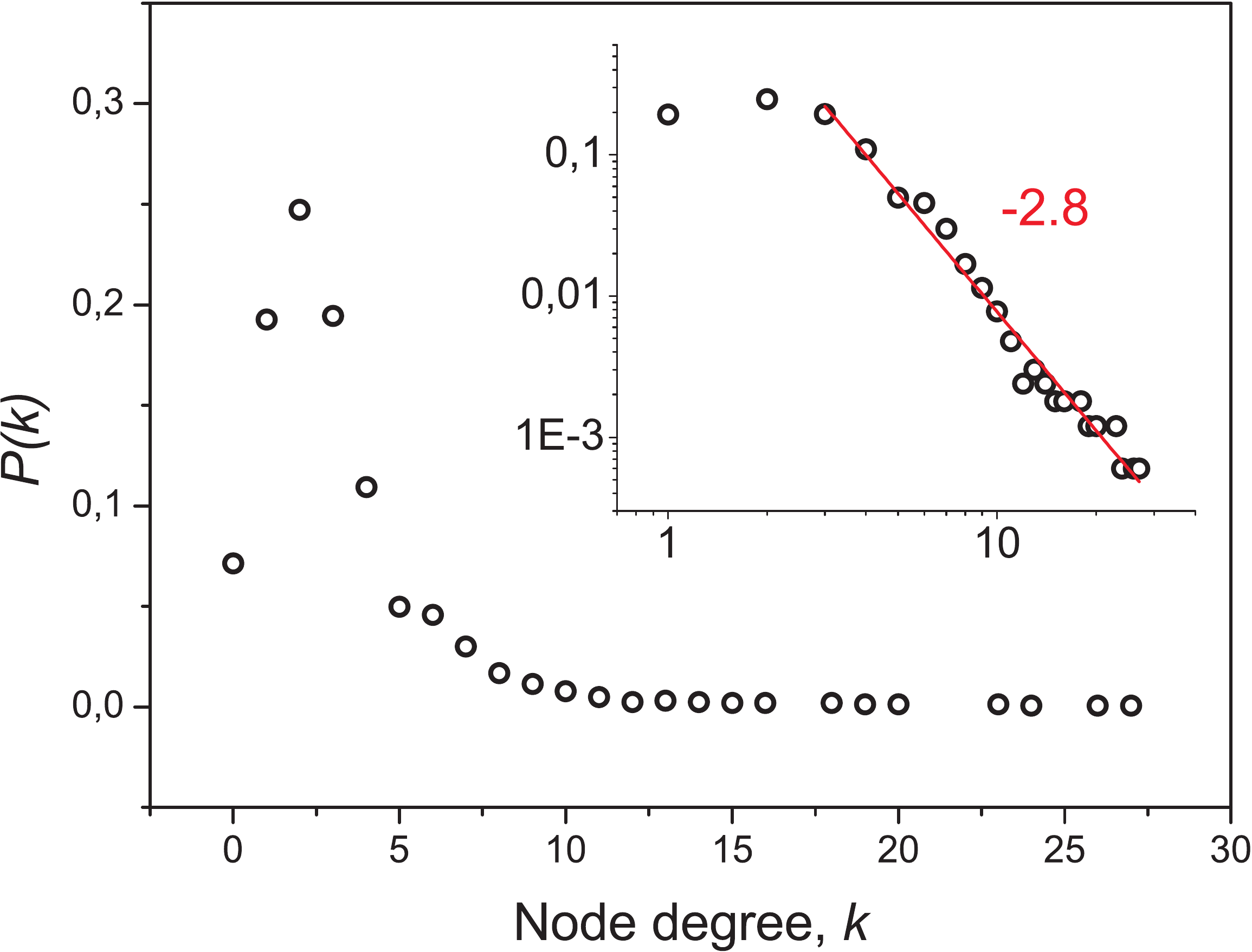}%
\caption{%
(Colour online) The node degree distribution $P(k)$ of the co-authorship network of the CMP journal. The inset contains the same plot in a double log scale: the linear fit for $k\geqslant 2$ is shown by red line with a corresponding slope marked.}
\label{fig_P(k)}
\end{figure}

The weighted co-authorship network is built to analyse the collaboration connections between CMP authors (see also \cite{Mryglod2007,networks,networks2,networks3}). Each of 1666 authors who has ever contributed to the journal is marked with a single node, while the link between the pair of the nodes means that at least one joint paper was published in CMP. The link strength (i.e., its weight) is proportional to the number of co-authored publications. The resulting network is quite sparse: it is characterized by a low value of density: $d\approx 0.002$ which is defined as the percentage of all potential connections that are indeed presented in the network. This is a rather typical property of the so-called scale-free complex network, in which there are a lot of nodes having few links and few nodes having a lot of links (see also figure~\ref{fig_P(k)}). In particular, such a situation is typical of co-authorship networks. Considering the density level as the indicator of similarity between the nodes \cite{Scholz2015}, one can assume that the community of CMP authors is rather heterogeneous: supposedly,  in terms of scientific interests.

Collaboration networks also belong to a class of small-world networks. This implies a high tendency of the nodes to cluster together along with the short average distances between their pairs. The CMP co-authorship network is strongly connected indeed: the average value of Watts-Strogatz clustering coefficient $\langle C \rangle$ (see \cite{Watts98})  approaches  0.9 which exceeds the corresponding value for the random network of the same size by more than 450 times! The average distance between the pair of reachable nodes is very close to the well-known ``six degrees of separation'' notion \cite{SixDegrees_1,SixDegrees_2}. The linear size of the entire network is characterized by a maximal shortest path --- \emph{diameter}, which is equal to 13 for CMP.
 High level of the network connectivity despite its sparseness is natural for small-world networks, since the clustering coefficient shows a local tendency of the node's neighbours to interconnect over the network --- it works for each connected component separately~\cite{Watts98}.

\begin{table}[!t] 
\caption{The numerical characteristics of
the co-authorship network of the CMP journal. $N$: the number of
nodes; $L$: the number of links; $\langle k \rangle$,
$k_{\mathrm{max}}$: the mean and maximal node degree,
respectively; $\langle C \rangle$: the mean clustering coefficient; $\langle l \rangle$,
$l_{\mathrm{max}}$: the mean and maximal shortest path length; $N_{\mathrm{LCC}}$, $N_{\mathrm{nLCC}}$: the size of the largest (LCC) and the next-largest (nLCC) connected components, respectively; $N_{\mathrm{i}}$: the number of isolated nodes.}%
\vspace{2ex}
\centering
 \label{tab_Coauth_numbers}
 \small
\begin{tabular}{|c|c|c|c|c|c|c|c|c|c|c|}
\hline\hline Parameter&$N$&$L$&$k_{\mathrm{max}}$&$\langle
k\rangle$&$\langle C \rangle$&$\langle
l\rangle$&$l_{\mathrm{max}}$&$N_{\mathrm{LCC}}$&$N_{\mathrm{nLCC}} $&$N_{\mathrm{i}}$\\\hline Value&
1666&2598&27&3.1&0.87&5.6&13&437 (26.2\%)&43 (2.6\%)& 119 (7.1\%)
\\ \hline\hline
\end{tabular}
\end{table}

The distribution of the node degrees (the number of links connected to a node) is shown in figure~\ref{fig_P(k)}: its tail can be described by power law  $P(k) \sim k^{-2.8}$. One can clearly see that most frequently the authors have two collaborators within CMP, while the corresponding average value is equal to 3.1 --- this also points to the theoretical character of the research results published in CMP~\cite{Newman2001,Newman2004}.
Five nodes are characterized by a degree greater than 20 representing the most collaborative authors:  I.V.~Stasyuk (27), R.R.~Levitskii (26), M.V.~Tokarchuk (24), M.F.~Holovko (23) and Yu.M.~Vysochanskii (23). A significant overlap with the list of the most productive authors is quite natural. These and other most important numerical characteristics of the network are summarized in table~\ref{tab_Coauth_numbers} (the definitions of typical network properties can be found, for example, in \cite{networks,networks2,networks3}).

\begin{figure}[!t]
\centerline{\includegraphics[width=0.9\textwidth]{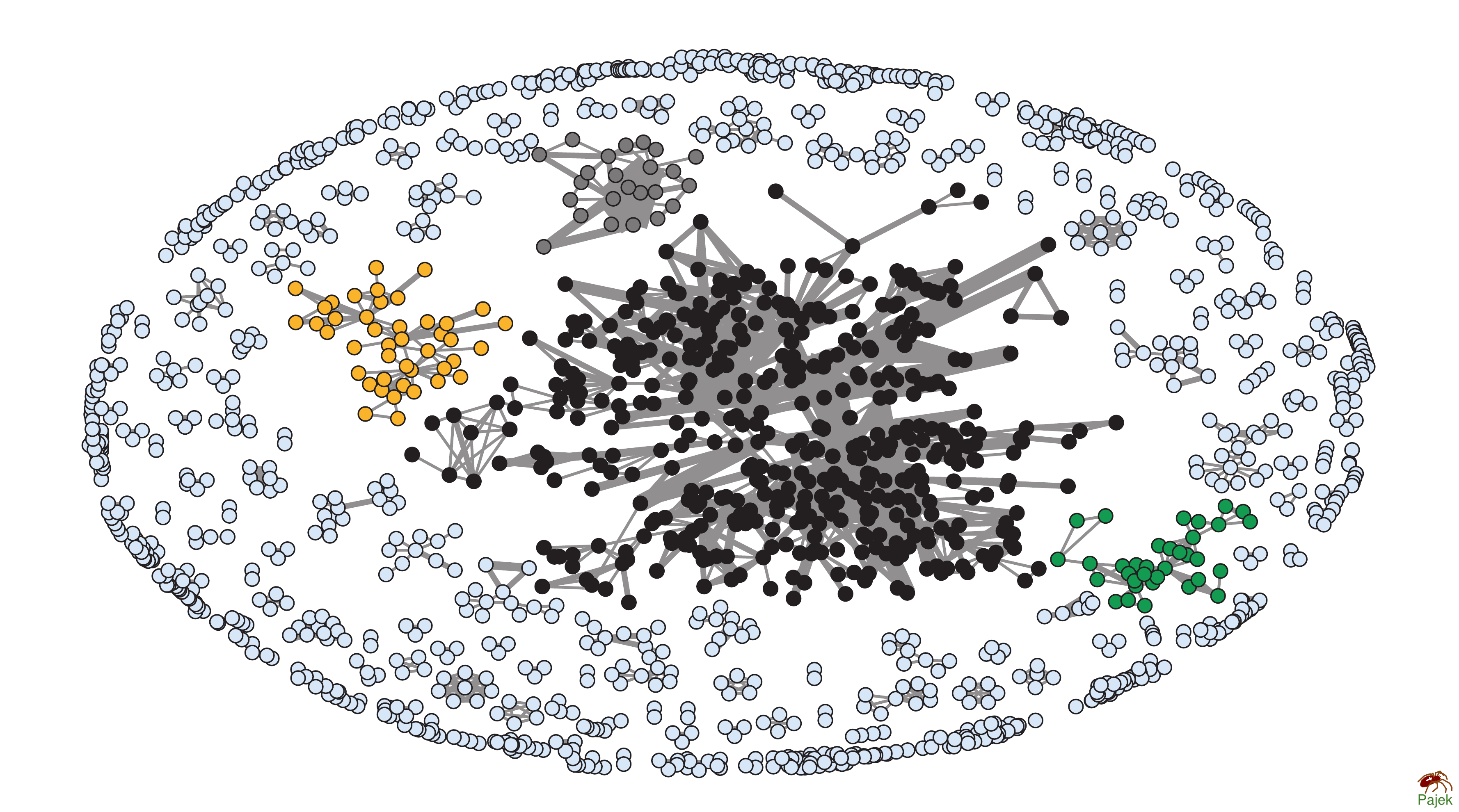}}
\caption{(Colour online) The weighted co-authorship network of CMP journal (1993--2017). Four largest connected components containing more than 20 nodes are highlighted by different colours: black --- 437~nodes~(LCC), orange --- 43 (nLCC); green --- 34; grey --- 24. This and further networks are generated using Pajek network visualization software~\cite{Pajek}.}
\label{fig_CoauthNetw_all_components}
\end{figure}

Structural properties of the network can be taken into account for the component analysis. To this end, 387 separate connected components are distinguished: 119 of them contain a single (i.e., isolated) node representing the authors who contributed to CMP journal without co-authors. On the other hand, more than 26\% authors belong to the largest connected components which is approximately ten times bigger than the next-largest one, see figure~\ref{fig_CoauthNetw_all_components}.

A structure with clearly defined largest connected components is typical of co-authorship networks as well as of numerous other real networks. Such a component whose size extensively increases with the network growth is called Giant Connected Component (GCC). It usually contains more than a half of all the nodes for collaboration networks \cite{Newman2001,Barabasi2002}. Meanwhile, the defined largest group of CMP network joints together only approximately 1/4 of all the journal contributors. Let us call it here the Largest Connected Component (LCC). Still, it can be considered as a ``core'' of journals' authors that actively submit manuscripts, collaborate within the subject field and, thus, form an expert environment required to provide a proper peer-reviewing process. The notions of centrality can be used to find the most influential authors within the LCC. The \emph{degree centrality} is defined by the number of links connected to a node. The most collaborative authors listed above are in TOP5 according to this measure. The \emph{closeness centrality} is a measure of closeness to any node within the connected fragment of the network: its high value marks the authors that can be referred to as ``good spreaders'' of information within the network. These authors are comparatively close to the numerous other authors and, therefore, they are informed about the experts in different subareas. The authors characterized by a high closeness centrality can supposedly play the role of the editors to distribute the manuscript between the journal sections. The following names are in TOP10 list according to this measure: I.M.~Mryglod, M.V.~Tokarchuk, A.D.~Trokhymchuk, I.P.~Omelyan, I.R.~Yukhnovskii, Ja.M.~Ilnytskyi, M.F.~Holovko, E.M.~Sovyak, M.P.~Kozlovskii, Yu.~Holovatch. Another centrality measure is \emph{betweenness}, which is proportional to the number of the shortest paths between the pairs of other nodes that go through the node or the link. The nodes characterized by the hight value of betweenness occupy the crucial positions in the network being a bridge connecting its separate parts. The corresponding authors might be better informed about the expert environment within the neighbouring subject area or a distant collaborative group. This can be sometimes useful in order to find peer-reviewers. The TOP10 list according to the betweenness values contains the following names: R.R.~Levitskii, A.D.~Trokhymchuk, Ja.M.~Ilnytskyi, M.V.~Tokarchuk, I.M.~Mryglod, I.R.~Yukhnovskii, M.P.~Kozlovskii, T.~Krokhmalskii, M.F.~Holovko, Yu.~Holovatch. All centrality measures indicate the importance of the authors in information spreading due to numerous links or an exceptional position in the network. It can be seen that some authors can be considered as central in different senses.

Certainly, the remaining part of the network should not be neglected: a separate group of the authors can represent a topical area more distant from the usual journal's field or another collaborative cluster. The most central authors for the next-largest connected component (nLCC) are: A.G.~Zagorodny, B.I.~Lev, V.M.~Loktev and  P.M.~Tomchuk (the component consisting of 43 nodes);   L.A.~Bulavin, A.V.~Chalyi and N.I.~Lebovka (34 nodes); O.M.~Voitsekhivska, M.V.~Tkach, A.M.~Makhanets and Ju.O.~Seti (24 nodes). Their central positions are indicated by the highest values of both closeness and betweenness: all the authors listed above are in TOP5 according to both centrality measures.

Two different approaches can be used to investigate the dynamics of the network structure. On the one hand, the cumulative change of the network properties shows the way it gradually grows and the way its structure becomes more and more ``mature''. In this case, all the available data since the very start are taken into account. On the other hand, the so-called ``sliding window technique'' is useful to compare different periods of the journal's life. To this end, only the data relevant to the fixed time periods (i.e., time windows) are used to build the sequence of networks and to compare their properties afterwards. Both approaches enable one to get to know about the journal's authors. While the first approach summarizes the collaboration events for the entire history, the second one reveals the active connections between the authors.

\begin{figure}[!t]
\centering
\includegraphics[width=0.48\textwidth]{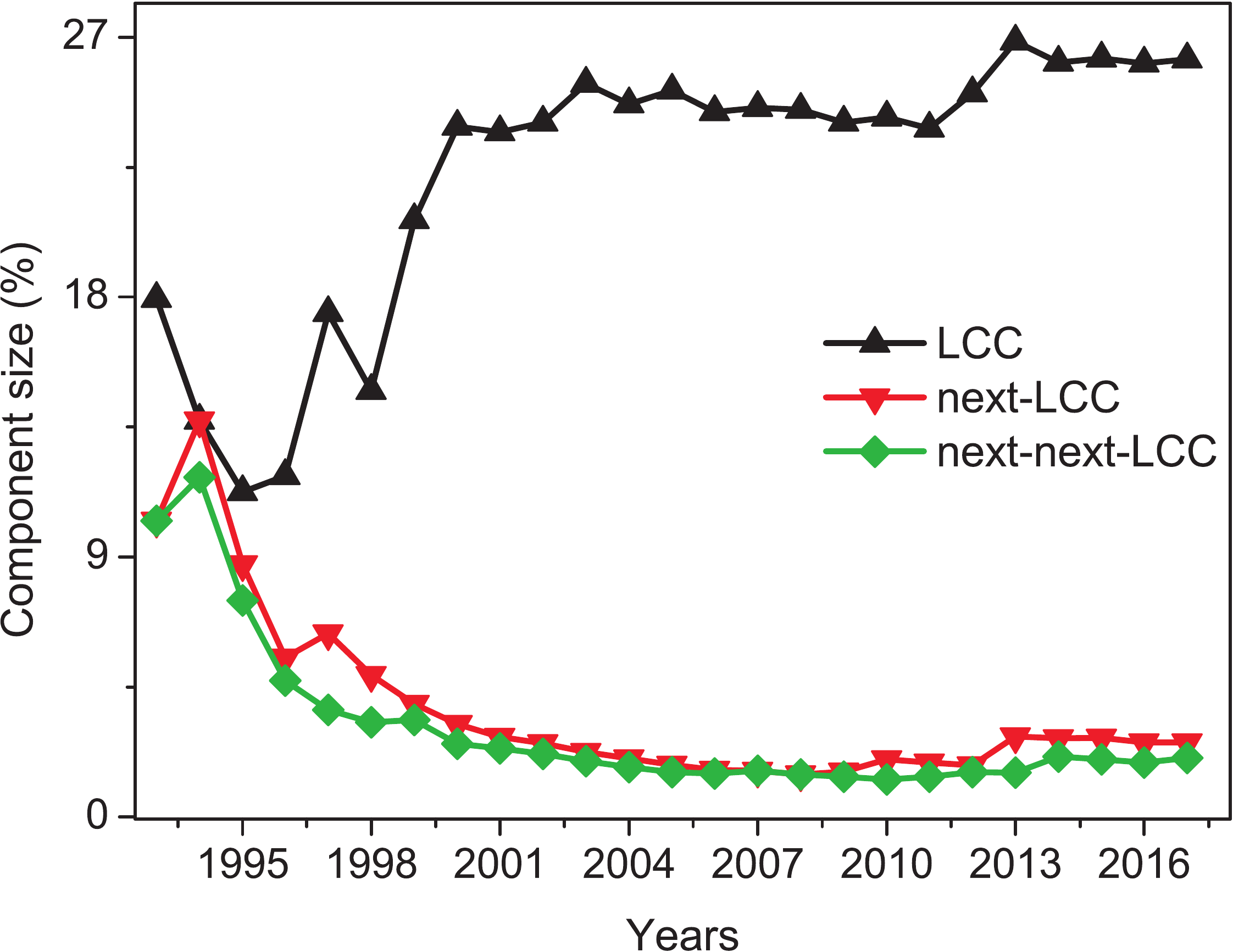}%
\caption{%
(Colour online) The annual change of the shares of the nodes included into three largest connected components of the CMP co-authorship network: cumulative growth is considered. LCC is the largest connected component; next-LCC and next-next-LCC are the next- and next-next-largest ones, respectively.}
\label{fig_LCCgrowth}%
\end{figure}

As concerns the issue of the network structure which could be considered as a ``mature'' structure for CMP, let us consider the cumulative growth of the network to define a kind of benchmark: the shares of nodes included into the LCC, the nLCC and the next-nLCC (the third largest one) versus time are shown in figure~\ref{fig_LCCgrowth}. The shape of the node degree distribution does not change significantly in time, while the largest connected component becomes more and more distinguishable: after the first few years, the network structure is more homogeneous: it contains several competitive groups of authors. However, since 2001--2002, the difference between LCC and the next-largest components becomes noticeable.

The evolution of the co-authorship network structure starts at the point when there exists a number of the connected fragments along with isolated nodes. Some connected groups grow by adding new nodes while new data are taken into account. Another scenario is combining two or more connected fragments together. Usually, after such a merge, the largest connected component becomes defined: it gradually grows while the next-size components remain comparatively small. During the first three years of CMP history, several connected groups of authors competed for the LCC title. Actually, the component containing the following 7 authors became the source of future LCC: I.R.~Yukhnovskii, S.I.~Sorokov, A.P.~Moina, R.R.~Levitskii, V.O.~Kolomiets, I.M.~Idzyk and O.V.~Derzhko. After gradually adding a few authors, the first merge happened in 1996 when a triple consisting of A.M.~Shvaika, Yu.V.~Sizonenko and I.V.~Stasyuk entered the largest fragment. However, a crucial merge was in 1997, when two smaller components complemented LCC:  M.F.~Holovko, A.F.~Kovalenko, O.P.~Antonevych, E.M.~Sovyak, I.A.~Protsykevich, I.M.~Mryglod, A.M.~Hachkevych, R.I.~Zhelem, O.E.~Kobryn, V.G.~Morozov, M.V.~Tokarchuk and I.P.~Omelyan joined. Starting from that point, the difference between the LCC and the next-LCC was increasing. Therefore, it took about 7 years for CMP (1993--2000) to get its current co-authorship structure.

As far as the period of seven years turned out to be sufficient for CMP to form its inherent co-authorship structure, a comparison of two periods (1993--2005 and 2006--2017) seems reasonable. Moreover, starting from 2006, CMP data were used to calculate the Journal Impact Factor. Some other changes  aimed at improving the CMP's visibility were adopted. For example, certain changes in editorial processing of the manuscripts submitted to CMP were introduced. Collaboration with the CrossRef system that started in 2011 provided DOI numbers for all publications. Synchronization of all the papers with the e-Print archive arXiv of an Open Archives Initiative (www.arxiv.org) was organized in 2011 as well.

The shapes of the node degree distribution $P(k)$ as well as the distributions of the component sizes are very close for co-authorship networks built during these two periods. It can be seen in table~\ref{tab_Coauth_numbers_02} that their numerical characteristics are similar as well. The only exception is the size of LCC --- it contains a smaller part of nodes for the second period. A possible explanation of a more fragmented network structure is that after the CMP became properly promoted and more visible, it attracted new authors. An increased number of nodes $N$ for 2006--2017 period supports this version as well as a decreased number of papers by the authors from ICMP does (see figure~\ref{fig_icmp_papers}). Another evidence is a larger number of countries represented in CMP during the second period (see the next section).
\begin{table}[!t] \caption{Numerical characteristics of
the co-authorship network of the CMP journal. Notations are the same as in table~\ref{tab_Coauth_numbers}.}%
\vspace{2ex}
\begin{center}
 \label{tab_Coauth_numbers_02}
 \small
\begin{tabular}{|c|c|c|c|c|c|c|c|c|c|c|}
\hline\hline Period&$N$&$L$&$k_{\mathrm{max}}$&$\langle
k\rangle$&$\langle C \rangle$&$\langle
l\rangle$&$l_{\mathrm{max}}$&$N_{\mathrm{LCC}}$&$N_{\mathrm{nLCC}} $&$N_{\mathrm{i}}$\\\hline\hline
1993--2005&775&1130&23&2.9&0.86&4.6&9&195 (25.2\%)&14 (1.8\%)& 59 (7.6\%)
\\\hline
2006--2017&1111&1589&15&2.86&0.89&4.73&13&179 (16.11\%)&24 (2.2\%)& 80 (7.2\%)
\\\hline\hline
\end{tabular}
\end{center}
\end{table}

\begin{figure}[!t]
\centerline{\includegraphics[width=0.85\textwidth]{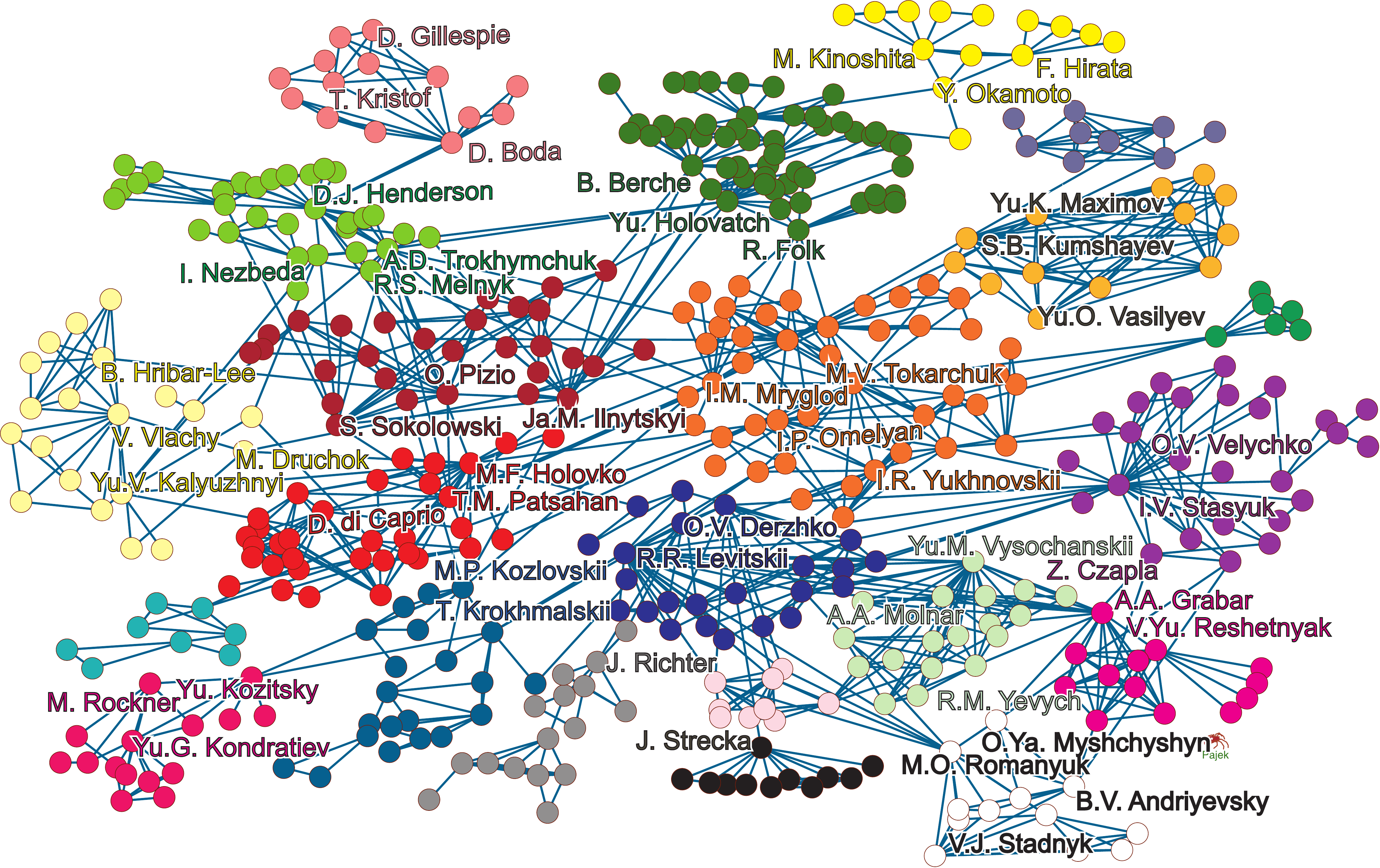}}
\caption{(Colour online) The largest connected component of CMP co-authorship network. Communities detected using the Louvain method are shown in different colours. The most central authors for communities containing more than 10 nodes are labelled by the corresponding  names of the authors.}
\label{fig_LCC_comm}
\end{figure}
32 authors are common for the LCC of the two networks considered. It is interesting to note that two authors that published the largest number of papers both during the first and the second periods, I.V.~Stasyuk and R.R.~Levitskii, are  members of the LCC during 1993--2005 but not of the LCC during 2006--2017. In the second network, these two authors are centered in the third-largest connected component consisting of 22 nodes.

The story about the co-authorship network structure is incomplete without mentioning the internal structure of its components. Heterogeneity of the links is usually observed \cite{Newman06}. This means that some groups of nodes are interconnected more tightly forming the so-called communities while the density of connections between such groups is lower. Numerous algorithms for community structure detections have been  suggested during the recent decade, see, e.g., \cite{Fortunato2009}, though each of them has its own drawbacks. Unfortunately, there is no a universal tool to perform an unambiguous division of nodes into  groups, but some are widely used to get the most probable result. The so-called Louvain method is used here to get one of the possible versions of the LCC community structure \cite{Louvain}. According to this method, 22 communities can be distinguished, see figure~\ref{fig_LCC_comm}. The nodes characterized simultaneously by the highest values of closeness, betweenness and degree centralities within each community (of the size greater than 10) are accompanied in the figure by the corresponding  names of the authors.

\section{International contribution to CMP}
\label{sec_geography}

\begin{figure}[!t]
\includegraphics[width=0.48\textwidth]{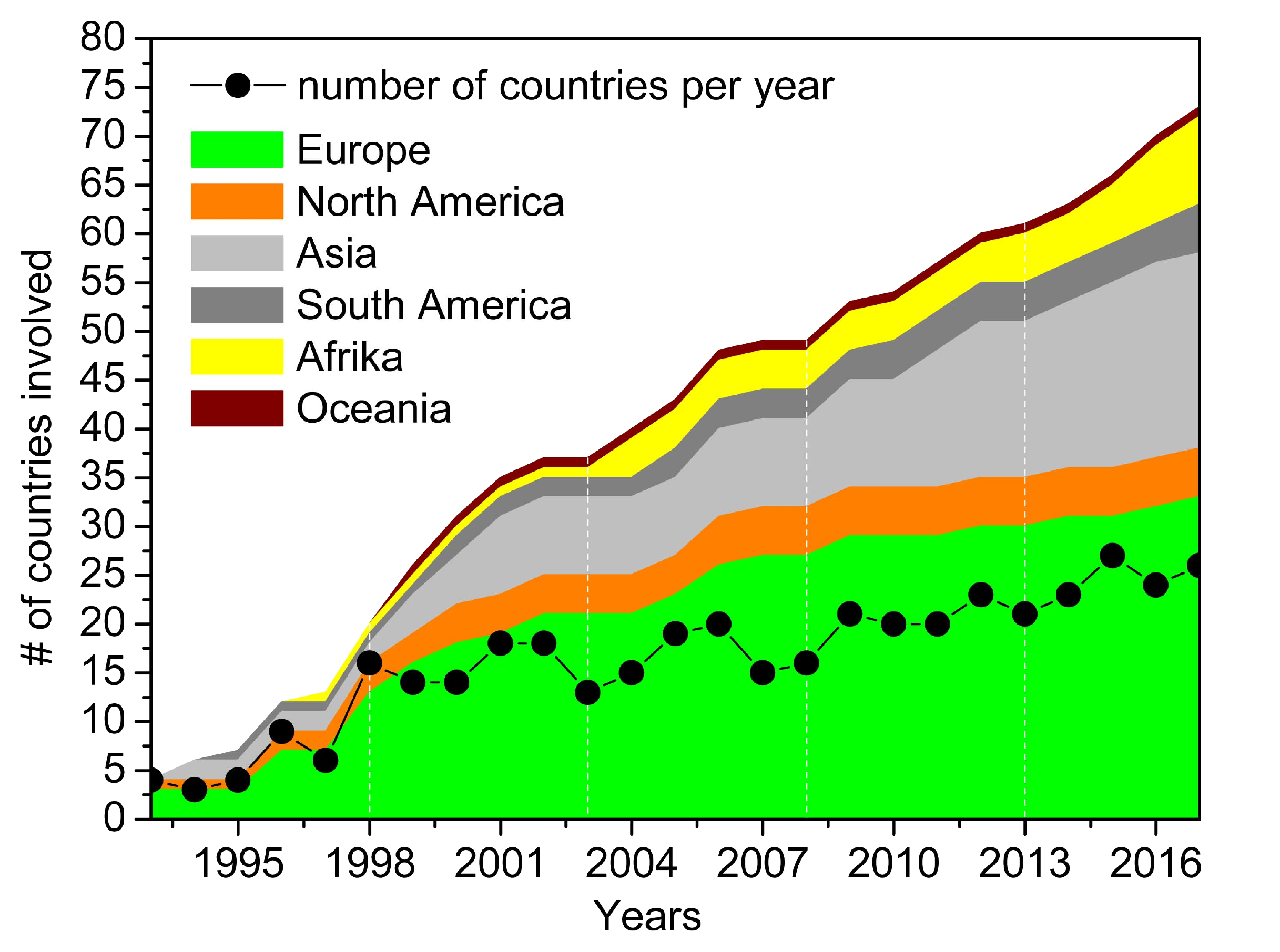}%
\hfill%
\includegraphics[width=0.48\textwidth]{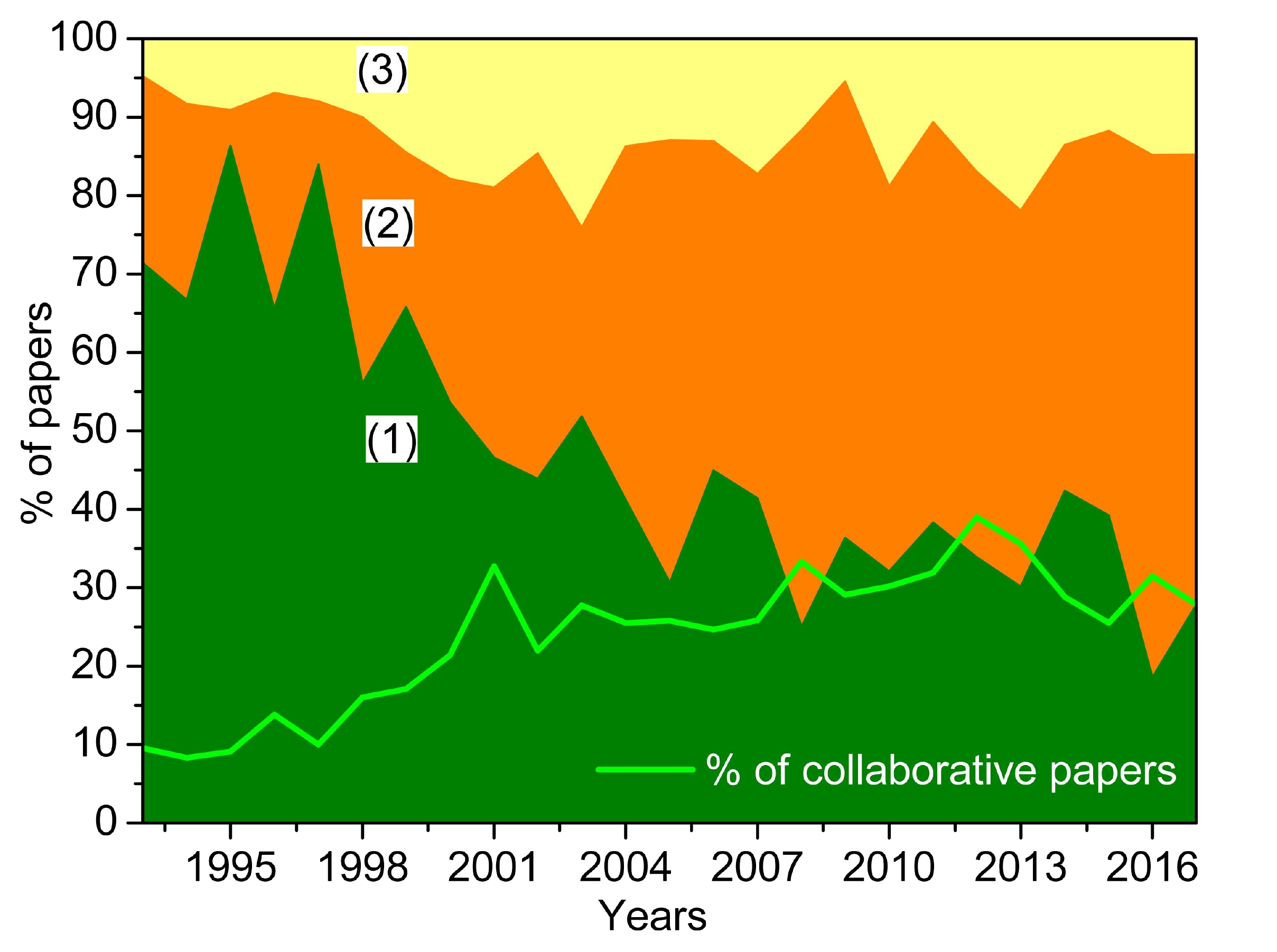}%
\\%
\parbox[t]{0.48\textwidth}{%
\caption{%
(Colour online) The dynamics of international collaboration in the CMP journal during 1993--2017: annual numbers of different countries per year are shown by symbols; cumulative numbers of countries from different parts of the world are shown by coloured areas.
}
\label{fig_Intern_growth}
}%
\hfill%
\parbox[t]{0.48\textwidth}{%
\caption{%
(Colour online) The percentage of the CMP papers authored by only Ukrainian authors (region 1), foreign authors (region 2) and written in collaboration of Ukraine and at least one foreign country (region 3). The line represents the percentage of the collaborative papers (two or more different countries are involved).
}
\label{fig_PercGeography}%
}%
\end{figure}

The affiliation data are used to investigate the geography of CMP journal. The spectrum of countries increases with the time passing, as it can be seen in figure~\ref{fig_Intern_growth}. In general, 72 foreign countries have  contributed into the journal between 1993 and 2017. TOP5 of the most productive foreign countries in terms of the number of CMP publications includes: Poland (117 relevant papers), Germany (115), USA (109), France (84) and Russia (68). In figure~\ref{fig_Intern_growth}, the cumulative numbers of countries from different parts of the world that scientifically contributed to CMP journal are shown for each year. A general tendency for increasing can be observed here and the distinguishable peaks supposedly correspond to the years when the collections of special papers were published. New countries have also enriched the international spectrum of CMP after its first Journal Impact Factor was announced in 2008. Of course, the increase cannot be infinite due to a limited number of countries.

Over 74\% of CMP papers are written by authors from one country, almost 21.5\% of them can be attributed to two countries and the rest 4.5\% for three or more countries. There is one paper, where the largest number of countries (6) are mentioned: 
\begin{itemize}
\item Palchykov~V., Kaski~K., Kertesz~J., \textit{Transmission of cultural traits in layered ego-centric networks}, Condens. Matter Phys., 2014, \textbf{17}, No.~3, 33802, \doi{10.5488/CMP.17.33802}.
\end{itemize}
44.4\% of all papers were published by Ukrainian authors only, and in  58\% of papers, Ukraine was mentioned at least for one author. Figure~\ref{fig_PercGeography} demonstrates how these shares change annually. It can be easily seen that the share of Ukrainian-only papers decreases, the percentage of joint papers of Ukrainian and foreign authors fluctuates around 15\% while the number of entirely foreign paper increases. On the one hand, this positive dynamics shows that CMP journal becomes more visible on the international scale. On the other hand, this follows the general tendency of science to become more collaborative (e.g., see \cite{2013Adams}). Figure~\ref{fig_PercGeography} also demonstrates the growth of an annual share of papers that can be attributed to two or more countries and thus can be called collaborative ones: the average value for the first half of the journal's history is equal to 20\% while it is 30\% for the second half.

\begin{figure}[!t]
\centerline{\includegraphics[width=0.95\textwidth]{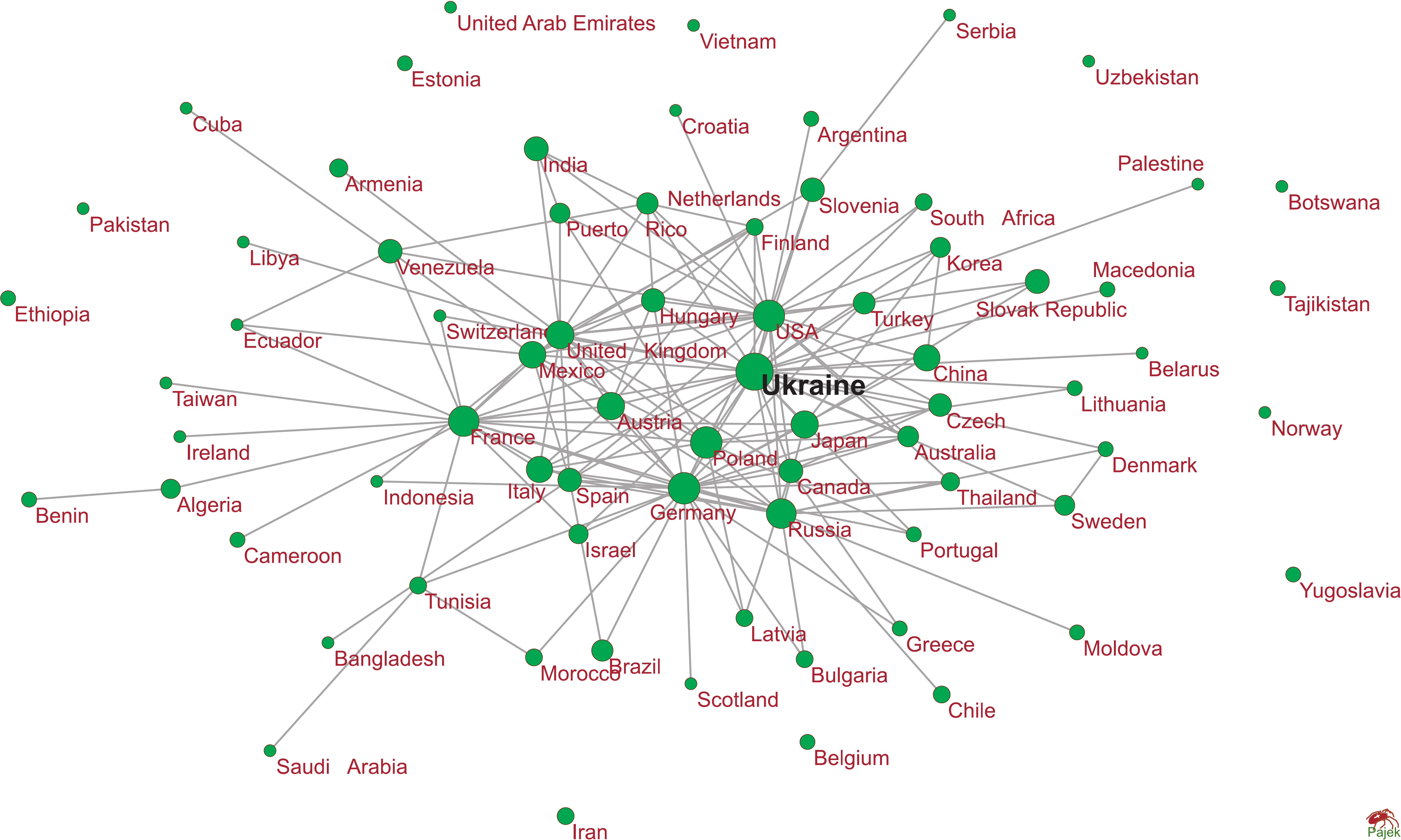}}
\caption{(Colour online) The co-authorship network at the level of countries: nodes represent countries, links mean that a pair of countries were listed in affiliation field of one CMP publication. The node size is proportional to the total number of papers written by the authors that mentioned the corresponding country in their affiliations.}
\label{fig_countr_coauthorship}
\end{figure}
Another kind of co-authorship network is built based on CMP data, see figure~\ref{fig_countr_coauthorship}. Here, nodes represent countries, while the link between the pair of them appears when two countries are mentioned in the affiliation field of at least one CMP publication. Besides Ukraine, which is linked with 31 other countries, the nodes of the highest values of degree representing the most collaborative countries correspond to Germany (28 collaborative countries), USA (25), France (20), UK (18) and Russia (15).

The most central positions in the network according to closeness and betweenness centralities are occupied by the same countries that are listed in the previous paragraph: Ukraine, Germany, France, USA, UK, and Russia. The majority of countries (83.6\%) belong to a single connected component --- Giant Connected Component, while the rest are represented by isolated nodes which means the absence of international collaboration of a certain country within CMP. The averaged node degree in the network is equal to 4.4, but within LCC it is $\langle k \rangle \approx 5.3$. Therefore, one can say that each country on average is connected with other 4--5 countries by co-authorship links in CMP journal. The average shortest path between pairs of connected nodes is rather short:  $\langle l \rangle \approx 2.4$, and the diameter is 5. Thus, the network is comparably compact.  Due to the fact that a large number of nodes are interconnected, the network is also characterized by a high value of the clustering coefficient: $\langle C \rangle \approx 0.7$. All the properties of the entire network  practically remain the same if calculated separately for two periods of CMP history. $\langle k\rangle \approx 3.3$, $\langle l\rangle\approx 2.3$, $l_{\mathrm{max}}=4$ and 81.4\% of nodes belong to LCC for the network based on the data for 1993--2005 (43 countries are in the network); $\langle k\rangle \approx 3.85$, $\langle l\rangle\approx 2.5$, $l_{\mathrm{max}}=5$ and 80.6\% of nodes belong to LCC for 2006--2017 period (68 nodes). As one can notice, only the diameter of the network is larger for the cumulated network.

\section{Topical spectrum of CMP}
\label{sec_PACS}
Since all CMP publications are labelled according to Physics and Astronomy Classification Scheme (PACS) \cite{PACS}, the corresponding numbers can be used in order to investigate the spectrum of scientific topics discussed in the journal.
The TOP10 of the most frequently used PACS in CMP is as follows: 05.50.+q ``Lattice theory and statistics (Ising, Potts, etc.)'' (mentioned in 54 papers); 61.20.Ja ``Computer simulation of liquid structure'' (54); 61.20.Gy ``Theory and models of liquid structure'' (48); 75.10.Hk ``Classical spin models'' (44); 77.80.Bh ``Phase transitions and Curie point'' (43); 05.20.Jj ``Statistical mechanics of classical fluids'' (40); 71.10.Fd ``Lattice fermion models (Hubbard model, etc.)'' (38); 61.20.Qg ``Structure of associated liquids: electrolytes, molten salts, etc.'' (38); 05.70.Jk ``Critical point phenomena'' (37); 64.60.Fr ``Equilibrium properties near critical points, critical exponents'' (35).
 PACS is an example of hierarchical index, where each next element denotes a more specific area of research. The first digits correspond to the most general classification denoting the most general disciplinary areas (of physics preferably). The part of the number up to the first dot indicates a subarea, and so on.
\begin{figure}[!t]
\centerline{\includegraphics[width=0.95\textwidth]{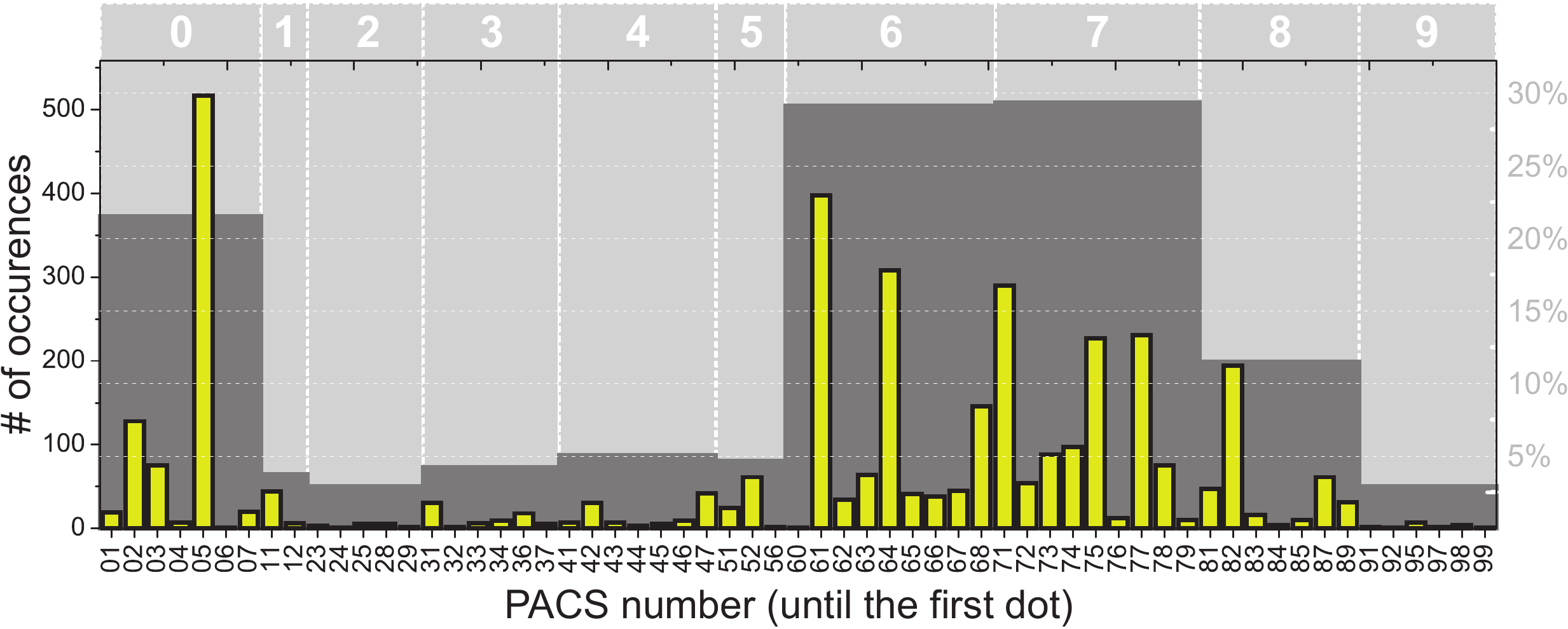}}
\caption{(Colour online) The frequency of using the PACS numbers in CMP papers: only its part up to the first dot is taken into account. The shaded areas show the share of indices that start from a certain digit (listed on the top).}
\label{fig_PACS_distr}
\end{figure}

The frequency of using the PACS numbers can give a hint about dominant topical directions of CMP journal, see figure~\ref{fig_PACS_distr}. Total numbers of the mentions of indices are presented here (one paper may be labelled by a few indices): the first part of PACS number up to the first dot is taken into account to build vertical yellow bars (the corresponding numbers are given below the horizontal axis); the shaded areas correspond to the percentage of all PACS indices starting with a certain digit (these are horizontally listed above the plot). It can be seen that two disciplinary areas are dominant: 6 (``Condensed Matter: Structural, Mechanical and Thermal Properties'') and 7 (``Condensed Matter: Electronic Structure, Electrical, Magnetic, and Optical Properties'') which corresponds to the title and the scope of CMP journal. The next topical direction --- 0 (``General'') ---  is followed by 8 (``Interdisciplinary Physics and Related Areas of Science and Technology''). Such a situation is  practically the same for both periods 1993--2005 and 2006--2017, except that the share of interdisciplinary research (code 8) in CMP increased from $\approx 6.6\%$ to $\approx 11.8\%$. The bars in the figure tell us a bit more about disciplinary subareas: it seems that primarily the application of statistical physics methods caused a majority of references to enter the category ``General'' (code 05) while  interdisciplinary connections mainly between physics and chemistry (code 82) are reflected in the ``Interdisciplinary Physics'' category. The most frequently used PACS in the interdisciplinary area are: 82.70.Dd ``Colloids'' and 82.35.Gh ``Polymers on surfaces; adhesion''. The following directions are the most popular within 6 and 7: ``Structure of solids and liquids; crystallography''; ``Equations of state, phase equilibria, and phase transitions''; ``Electronic structure of bulk materials''; ``Magnetic properties and materials''; ``Dielectrics, piezoelectrics, and ferroelectrics and their properties''.

\begin{figure}[!t]
\centerline{\includegraphics[width=0.5\textwidth]{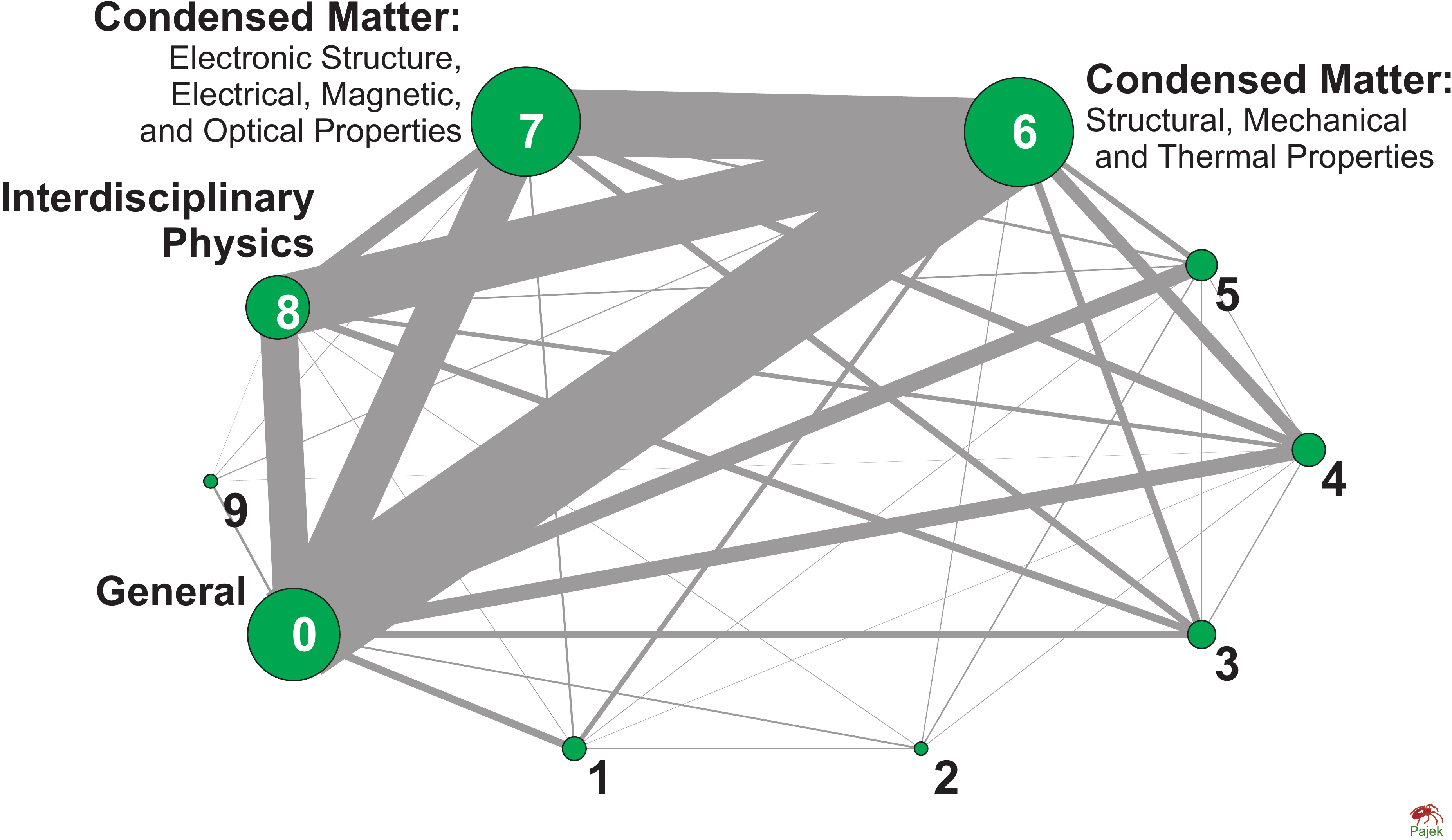}}
\caption{(Colour online) The weighted network of PACS numbers co-used in CMP papers: only the first digit of code is taken into account. Node size represents the number of papers, while the weight of a link relates to the number of papers, where two PACS number were mentioned.}
\label{fig_PACS_netw_1st_digit}
\end{figure}

In order to see in what way the dominant CMP topics are interconnected, another kind of network is constructed using the data on the PACS number assigned to each CMP paper. Here, a node represents PACS index or its part while a link connects the pair of indices if they co-occur in a paper. The link weight  then relates to the number of papers where two indices are both mentioned. Depending on the purpose of the investigation, the entire PACS number or just its part can be taken into account, representing the most particular topic or general subarea, correspondingly. The interrelations between the dominant disciplinary areas reflected in CMP are shown in figure~\ref{fig_PACS_netw_1st_digit}. One can see that the strongest connections between the leading areas lie along two triangles: 0--6--7 and 0--6--8. Obviously, the majority of CMP interdisciplinary research is probably related to chemical aspects of studying the properties of condensed matter, probably using statistical physics approaches, see also figure~\ref{fig_PACS_distr}.

\begin{figure}[!t]
\centerline{\includegraphics[width=0.95\textwidth]{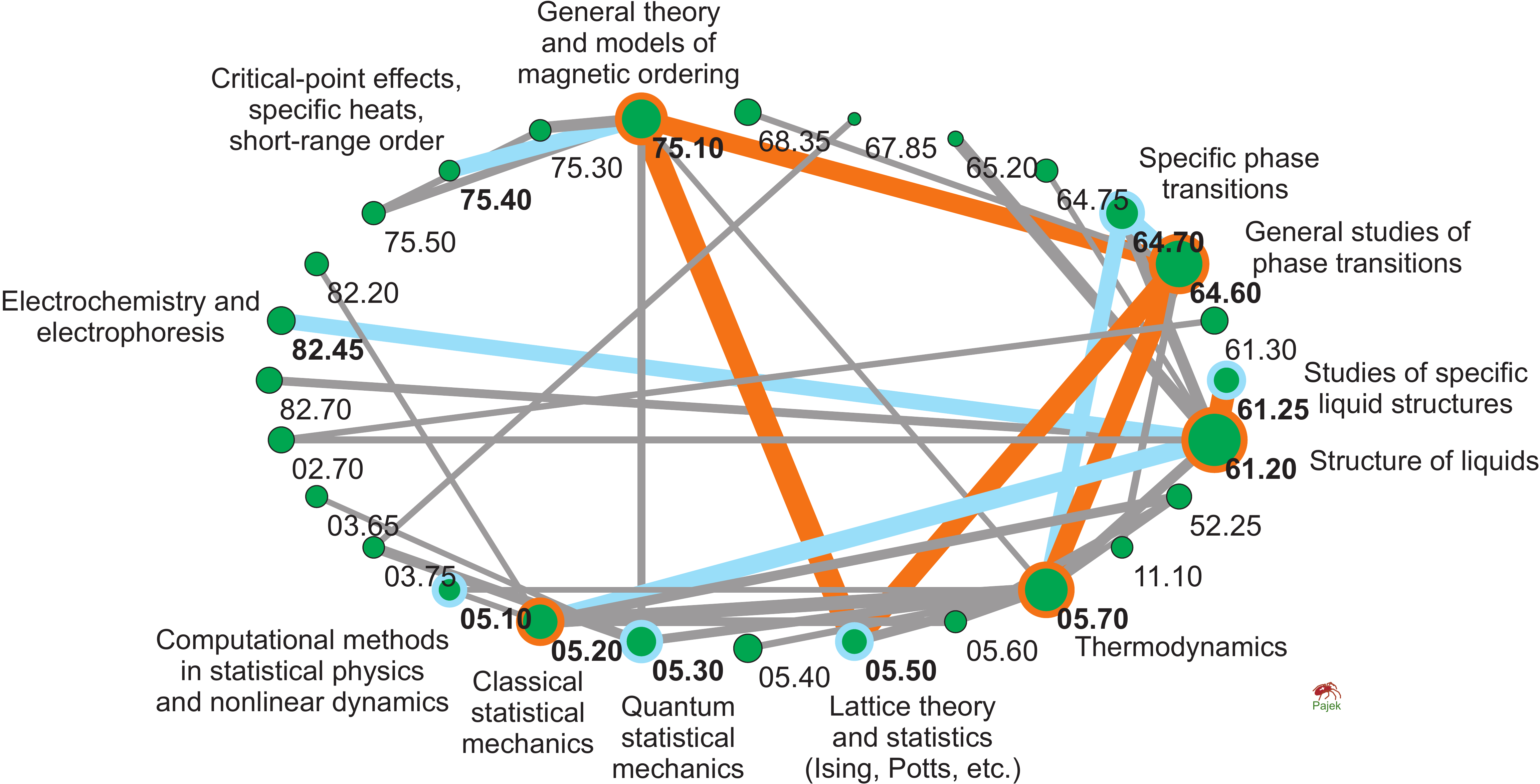}}
\caption{(Colour online) The largest connected components of the reduced network of PACS numbers co-used in CMP papers: only the part up to the second dot is taken into account. Minimal node weight (the number of papers) is 2 while minimal link weight (the number of joint papers) is 10. TOP10 nodes according to the number of papers and TOP10 links according to their weights are highlighted: first 5 positions in orange and next 5 positions in blue.}
\label{fig_PACS_netw_2nd_dot}
\end{figure}
Let us consider the network of co-used PACS numbers, taking into account its parts up to the second dot, in order to see the topical structure more in detail. Such a network contains 370 nodes and is characterized by a rather high connectivity: $\langle C \rangle \approx 0.6$, density is higher compared to its values for the networks considered before $d\approx 0.03$, and about 96\% of all nodes belong to the largest connected component. 123 nodes represent topical indices that are used only  once and thus can be considered as rare or occasional: the network becomes practically connected (except 2 nodes) after their removal. To reveal the most significant connections between the topics, let us reduce the network again. This time, all the links indicating less than 10 joint CMP papers for a pair of truncated PACS are removed. The threshold equal to 10 is chosen experimentally: the frequency of links decreases fast while their weights approaches  10 and much more slowly afterwards. The resulting networks are rather fragmented, but its LCC containing 28 nodes (11.3\%) can be considered as a topical core at the chosen level of accuracy. The most reliable connections between the most frequently used PACS number up to the second dot are presented in figure~\ref{fig_PACS_netw_2nd_dot}. It can be seen that the strongest triples typically contain different topical subareas that supposedly combine the method and the subject of research.

\section{Statistics of downloads vs. citations}
\label{sec_downloads}

Bibliographic information was used in previous sections to describe CMP journal, its authorship and different kinds of relations between publication data. However, the purpose of the next section is to describe the impact made by CMP publications. In order to describe an external ``impression'' made by the journal, other sources of data should be used. Of course, citation analysis is a commonly used way of quantifying such an impact considered as the most relevant to the content of publications. Citation-based metrics, characterizing the journal as a whole, are widely discussed and can be often obtained directly from scientometric resources such as Web of Science \cite{WoS} or Scopus \cite{Scopus}. On the other hand, another kind of data have been recently  considered as a source of information about the research papers. Discussions in social networks, posts in blogs, bookmarks and personal recommendations --- all these data coming from online activities of users tells us something new about scientific publications. The so-called \emph{altmetrics} can be computed using ``digital traces'' in the Internet characterizing resources (i.e., papers, journals) in a new way, which can be considered as complementary to traditional scientometrics \cite{Priem2014,Sud2014}. The so-called \emph{usage metrics} are closely related to them. They have been known since the times when librarians were trying to analyse the interests of readers, but  they have acquired additional meaning in the era of Internet. The statistics of pageviews, clicks or downloads manifest the users' interest to particular resources. An attractiveness rather than scientific value is reflected then because these actions can be made even before the content is read.

The download statistics of CMP provided by the publisher is incomplete and is not declared to be COUNTER compliant, see \cite{COUNTER}. Therefore, it should be used with caveats as a basis for comparison of different journals. However, it is a useful source of information about the attractiveness of CMP publications within the framework of a single journal. The database contains download records for full texts of all CMP papers starting from 2009 with one month resolution. Therefore, the publication window covers the entire CMP history while the download window is limited to the last 9 years (January~2009--December 2017). Thus, again 1282 research papers in total are available in the database. TOP5 list of the most downloaded CMP papers, according to the given download statistics, includes:
\begin{enumerate}
\item 	Imai T., \textit{Molecular theory of partial molar volume and its applications to biomolecular systems}, Condens. Matter Phys., 2007, \textbf{10}, No.~3, 343, \doi{10.5488/CMP.10.3.343}.
\item Gonchar N.S., \textit{Mathematical model of a stock market}, Condens. Matter Phys., 2000, \textbf{3}, No.~3, 461, \doi{10.5488/CMP.3.3.461}.
\item Hoover Wm.G., Hoover C.G., \textit{Nonequilibrium molecular dynamics}, Condens. Matter Phys., 2005, \textbf{8}, No.~2, 247, \doi{10.5488/CMP.8.2.247}.
\item Badiali J.-P., \textit{The concept of entropy. Relation between action and entropy}, Condens. Matter Phys., 2005, \textbf{8}, No.~4, 655, \doi{10.5488/CMP.8.4.655}.
\item Cabrera H., Marcano A., Castellanos Y., \textit{Absorption coefficient of nearly transparent liquids measured using thermal lens spectrometry}, Condens. Matter Phys., 2006, \textbf{9}, No.~2, 385,\\ \doi{10.5488/CMP.9.2.385}.
\end{enumerate}
The papers listed above were published either in special issues or in conference proceedings. One can only speculate about the reasons: do the features of interdisciplinary topic (papers 1 and 2) attract more attention? Which title sounds more intriguing: the general, ``review-like'' (papers 3 and 4) or, contrarily, more specific papers? These questions remain open so far. Besides, it looks quite natural that the most recent publications are not included into this list --- the older papers have a preference to collect downloads during a longer period even if the highest download rates are for the first period due to a novelty of the published results (one of the primary incentives to download \cite{Moed2005,Moed2016}).

\begin{figure}[!t]
\includegraphics[width=0.48\textwidth]{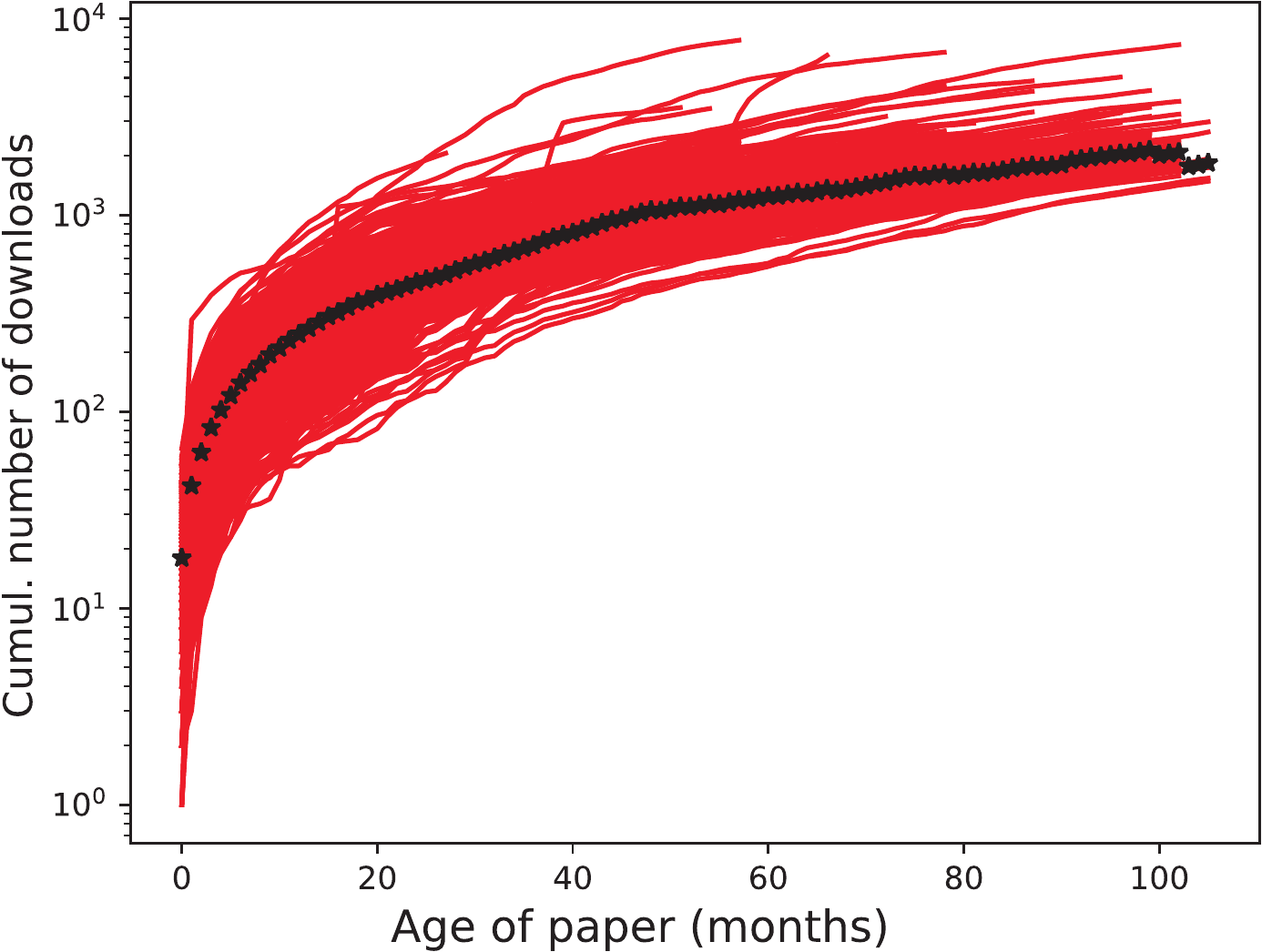}%
\hfill%
\includegraphics[width=0.48\textwidth]{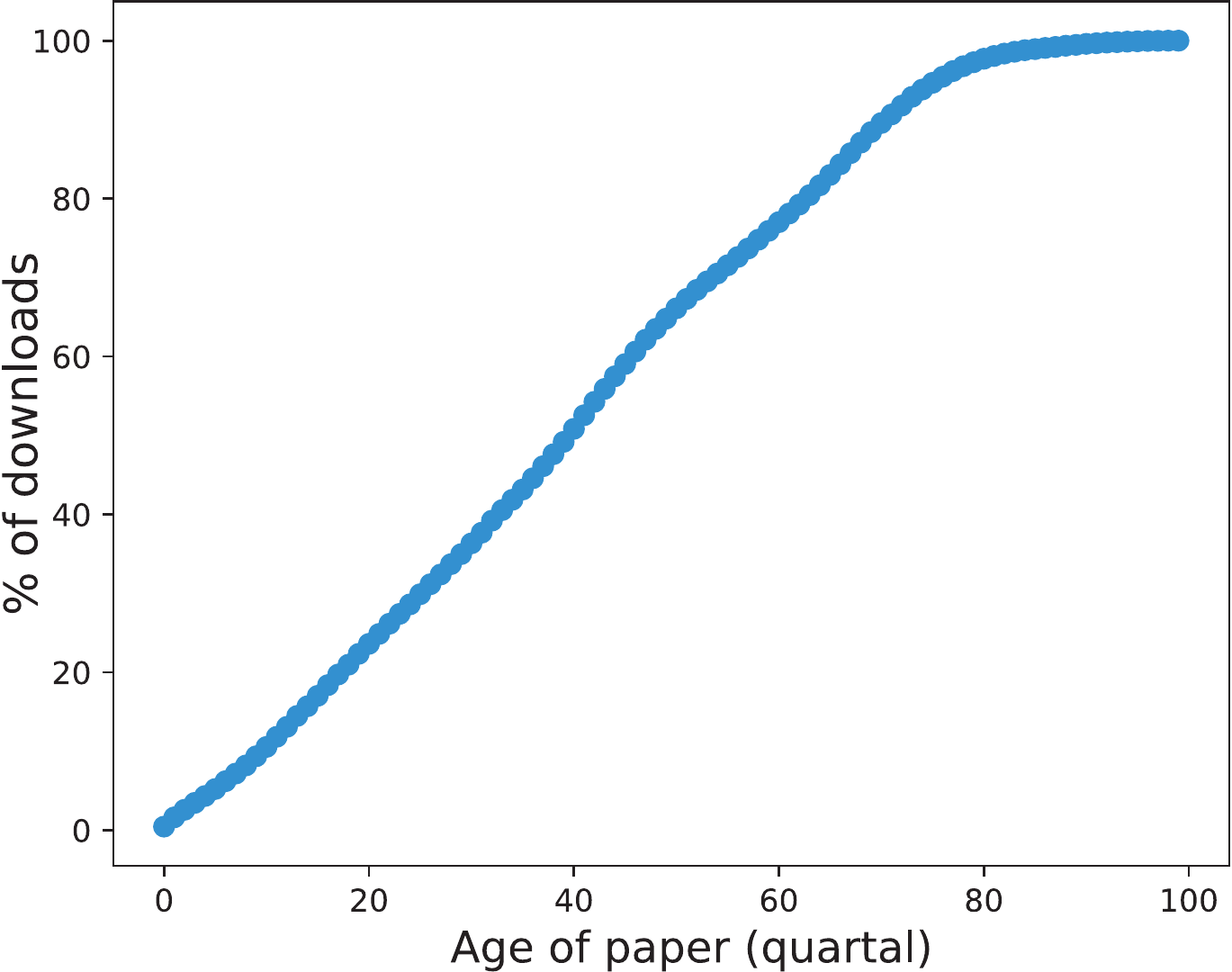}%
\\%
\parbox[t]{0.48\textwidth}{%
\caption{%
(Colour online) Cumulative number of downloads for
each paper individually vs. its age online. The median
values for papers of the same age are indicated by the symbols~$\ast$. Publication period: 2009--2017 (108 months in total).}
\label{fig_diach_all+bursty}
}%
\hfill%
\parbox[t]{0.48\textwidth}{%
\caption{%
(Colour online) Relative cumulative distributions of ages
of downloaded CMP papers. Time resolution is 3 months = 1 quarter. Publication period: 1993--2017 (100 quarters in total); observation period: 2015--2017. }
\label{fig_synchronous_2015-2017}%
}%
\end{figure}

The typical ``hottest'' period for collecting downloads for CMP papers can be detected in figure~\ref{fig_diach_all+bursty}: the most remarkable growth of cumulative downloads is observed for the first three months and later, after approximately 1.5 years, the downloading rate decreases. The so-called \emph{diachronous approach} was used to make this plot, i.e., the evolution of download process of each publication is investigated on its individual time scale starting from the moment of online publication. To compare individual processes, one should take only the papers with full download statistics. Therefore, only 514 papers that appeared online since 2009 are considered here. The median values calculated for the papers of the same age can be considered as the typical ones (denoted by $\ast$ in figure~\ref{fig_diach_all+bursty}) to define the characteristic curve describing the downloads accumulation process in CMP \cite{EPL}.

Another way of assessing the typical patterns of downloads for the entire journal is to use the so-called \emph{synchronous approach}, when the period of real time is chosen to analyse the statistics of downloads for papers of different age. All CMP papers published since its start in 1993 can be taken into account in this case. The plot in figure~\ref{fig_synchronous_2015-2017} is based on the cumulative data collected during the period 2015--2017. It tells us that 50\% of total downloads (vertical axis) for this period belong to the papers published during the previous 10 years (approximately 40 quarters on the horizontal axis). Therefore, one can conclude that CMP publications turn out to be attractive  during a comparatively long period of the paper's life.

Additional information about the online popularity of the journal can be provided by Google Analytics tool \cite{GA}. The latter is used to monitor the online visits to pages containing contents of CMP issues published since 2011 (the monitoring itself started from May 22, 2011). The traffic is analysed on IP basis to count users and to determine their approximate geographic locations. The obtained statistics cannot be used to investigate the readers' interests to different CMP papers, but it helps to get some general picture. For example, the rating of CMP issues can be generated\footnote{The Google Analytics results were accessed on May 5, 2018.} based on the number of users that visited its contents. Accordingly, the largest number of visits go to the first two regular issues (Nos. 1 and 2, 2011) published immediately after the collaboration with CrossRef system had started, i.e., since all CMP papers had got DOI numbers. The next positions in rating are almost fully occupied by special topical issues or the collections of conference publications.

\begin{figure}[!t]
\centerline{\includegraphics[width=0.93\textwidth]{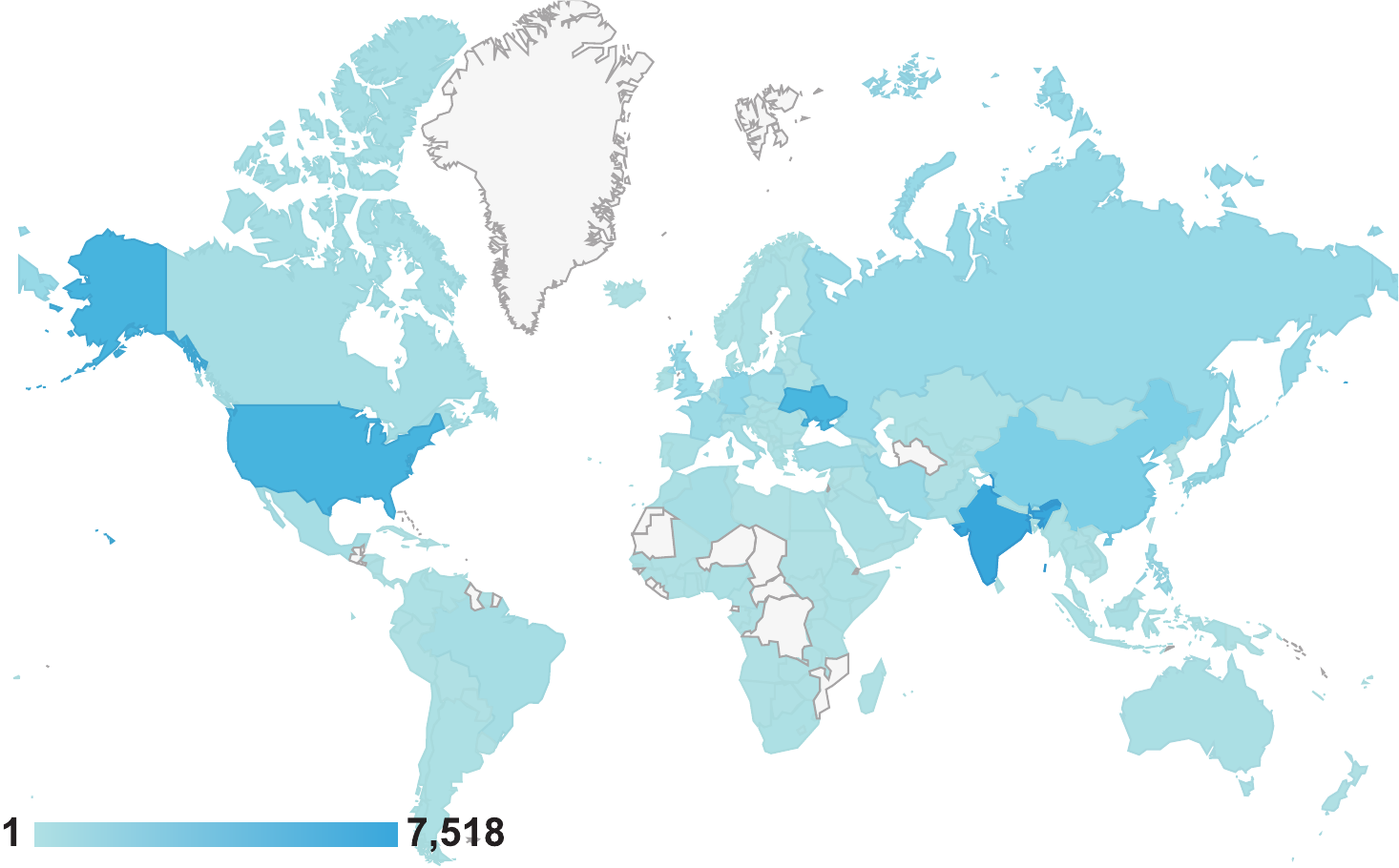}}
\caption{(Colour online) Geography of CMP website users based on Google Analytics data between May 22, 2017 and February 17, 2018.}
\label{fig_GA_map}
\end{figure}

Over 2\,000 pageviews and more than 600 users are reported on average per month by Google Analytics for CMP website. The approximate geography of users is shown on the map, see figure~\ref{fig_GA_map}. According to the reports, India, Ukraine and USA are three leaders by the number of interested users. The next positions in the corresponding TOP10 (by a descending value) are occupied by: China, Germany, Russia, UK, Japan, Iran and France.

\begin{figure}[!t]
%\vspace{3mm}
\centerline{\includegraphics[width=0.85\textwidth]{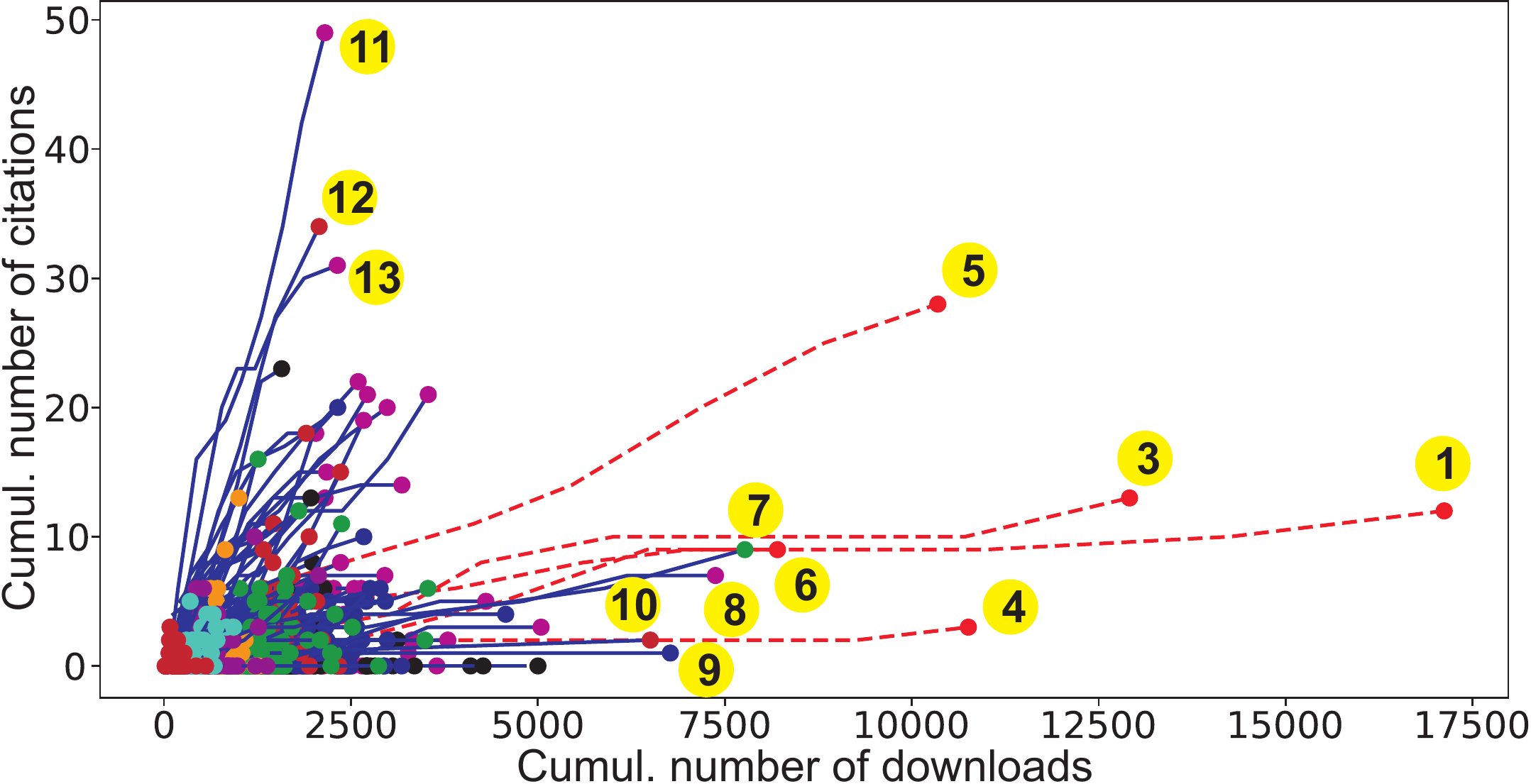}}
\caption{(Colour online) The dynamics of cumulative $(d,c)$ values for CMP papers published since 2009. The final cumulative values are marked with circles. Different colours denote different publication years. Five papers published within the period [2005--2009] are shown by dashed lines: they are characterized by cumulative downloads values that exceed the maximal ones for the rest of the papers. The papers labelled by numbers are mentioned in the text.}
\label{fig_d_vs_c_all}
\end{figure}

One of the hottest topics of interest is connected with correlation between downloads and citations. It is already known that both processes have some similarities as well as dissimilarities \cite{Moed2005,Moed2016}. Therefore, it is interesting to compare not only the cumulative values but also the evolution of both processes. The statistics of citations is acquired from Scopus database\footnote{Accessed February 11, 2018.}: 802 CMP papers published between 2005 and 2017 can be found in total, while 551 of them were cited at least once within the same period.

To investigate the correlations between $d$ (downloads) and $c$ (citations), an overlap between two datasets should be found. Only 591 CMP papers were published in 2009 or later, and thus are characterized by entire download and citations histories. These papers, therefore, are considered further. 320 of them were cited at least once according to Scopus database. The dynamics of cumulative $(d,c)$ values for each publication are depicted in figure~\ref{fig_d_vs_c_all}: the final cumulative values $(d,c)$ are shown by circles. Five papers published before 2009 were added into the figure (dashed lines) since their cumulative download scores, even without the known download statistics for several years after their publication, were still higher than the maximal values for the rest papers in figure. The most downloaded publications listed above can be found among these five papers (except paper No.~2 because it was published before 2005); the last one is ``Dielectric constant of the polarizable dipolar hard sphere fluid studied by Monte Carlo simulation and theories'' by Valisk\'{o}~M. and Boda~D. [Condens. Matter Phys., 2004, \textbf{8}, No.~2, 357, \doi{10.5488/CMP.8.2.357}] --- let us assign No.~6 to it. Among the other papers in figure~\ref{fig_d_vs_c_all}, the highly-downloaded but less cited ones can be found:
\begin{enumerate}
\item[7.] Brics M., Kaupu\v{z}s J., Mahnke R., \textit{How to solve Fokker-Planck equation treating mixed eigenvalue spectrum?}, Condens. Matter Phys., 2013, \textbf{16}, 13002, \doi{10.5488/CMP.16.13002}. 
\item[8.] Henderson D., \textit{Analytic methods for the Percus-Yevick hard sphere correlation functions}, Condens. Matter Phys., 2009, \textbf{12}, No.~2, 127, \doi{10.5488/CMP.12.2.127}. 
\item[9.] Sliusarenko O.Yu., \textit{Generalized Fokker-Planck equation and its solution for linear non-Markovian Gaussian systems}, Condens. Matter Phys., 2011, \textbf{14}, 23002, \doi{10.5488/CMP.14.23002}. 
\item[10.] Kalyuzhnyi Yu.V., Hlushak S., Cummings P.T., \textit{Liquid-gas phase behavior of polydisperse dipolar hard-sphere fluid: Extended thermodynamic perturbation theory for central force associating potential}, Condens. Matter Phys., 2012, \textbf{15}, 23605, \doi{10.5488/CMP.15.23605}.
\end{enumerate}

The opposite ``wing'' of the plot is supposed to describe the highly cited but less downloaded papers:
\begin{enumerate}
\item[11.]\v{C}anov\'{a} L., Stre\v{c}ka J., Lu\v{c}ivjansk\'{y} T., \textit{Exact solution of the mixed spin-1/2 and spin-S Ising-Heisenberg diamond chain}, Condens. Matter Phys., 2009, \textbf{12}, No.~3, 353,\\ \doi{10.5488/CMP.12.3.353}.
\item[12.] Ivaneyko D., Toshchevikov V., Saphiannikova M., Heinrich G., \textit{Effects of particle distribution on mechanical properties of magneto-sensitive elastomers in a homogeneous magnetic field}, Condens. Matter Phys., 2012, \textbf{15}, 33601, \doi{10.5488/CMP.15.33601}.
\item[13.] Shanker J., Singh B.P., Jitendra K., \textit{Extreme compression behaviour of higher derivative properties of solids based on the generalized Rydberg equation of state}, Condens. Matter Phys., 2009, \textbf{12}, No.~2, 205, \doi{10.5488/CMP.12.2.205}.
\end{enumerate}
Incompleteness of information about the downloads does not permit us to include into this list the following publications characterized by high citation scores:
\begin{itemize}
\item Mason T.G., Graves S.M., Wilking J.N., Lin M.Y., \textit{Extreme emulsification: formation and structure of nanoemulsions}, Condens. Matter Phys., 2006, \textbf{9}, No.~1, 193, \doi{10.5488/CMP.9.1.193}. 
\item  Shanker	J., Singh B.P., Jitendra K., \textit{Analysis of thermal expansivity of solids at extreme compression}, Condens. Matter Phys., 2008, \textbf{11}, No.~4, 681, \doi{10.5488/CMP.11.4.681}. 
 \end{itemize}

\section{Conclusions}
The purpose of this paper was to make a ``multidimensional'' quantitative description of a scientific journal --- Condensed Matter Physics. Different kinds of data were used to characterize the journal from various points of view. Since the database is limited due to natural reasons, this work is considered as a case study rather than some fundamental work. However, a number of features typical of other journals can be observed already on such a small scale. First of all, the co-authorship of CMP was analyzed using one of the most classical examples of social networks --- collaboration network of the journal's authors. Some cumulative as well as dynamical properties of this network were discussed. It already demonstrates the features of a small-world network: an average path between any two authors is approximately equal to 6 while the total number of them is over 1.5 thousand, and the maximal distance is just two times longer. These parameters tell us that information spreading can be very effective within the network of CMP authors. This is due to a tendency of the nodes to cluster together forming  expert groups. The analysis of the network structure allows one to find such collaboration groups as well as the links serving as bridges between them. All these features observed for the co-authorship network of CMP are also studied in dynamics. It is shown that approximately 7 years was sufficient to get a typical structure.

Another co-authorship network was built in order to investigate the international contributions to CMP journals. The collaborative relations between  72 foreign countries plus Ukraine were analysed. The network is characterized by a single connected component which interconnects more than 80\% of the nodes. Naturally, the network is rather egocentric since Ukraine plays the most significant and central role. Although the majority of CMP publications are related to Ukraine, the share of collaborative papers involving two or more countries increases. The total number of international links grows as well. The ratings of the countries that most actively contribute to CMP are built: the total number of papers as well as the level of collaboration are taken into account.

The topical spectrum of a scientific journal can be analysed in different ways. Here, the analysis of topical indices used by the authors in the papers is performed. The frequency analysis of PACS numbers taking their different parts into account enabled us to define the dominant subject areas as well as more scrupulously define the specific topical directions. The network of PACS connected by common papers contains a large number of nodes. Therefore, to consider the connections between topics at a desired level of detailing but avoiding an overload with ``rare'' or occasional indices, two thresholds were used to reduce the network. Only the nodes representing PACS numbers (their part up to the second dot is taken into account) occurring at least twice in CMP papers are considered. The minimal accepted link weight is 10. This makes it possible to concentrate on the most significant intertopical relations.

Finally, the data from external sources were used to analyse the impact made by CMP publications. Their attractiveness to online users was studied using the statistics of downloads of full texts of the papers. The process of downloads ageing is described --- i.e., reducing the intentions of users to download older publications. It is shown that the ``hottest'' period for downloads is the first three months after publication while after 1.5 years, the process of downloads becomes remarkably less intensive. The cumulative plot of downloads versus the papers' age demonstrates the pattern typical of CMP: not only the newest publications are of interest to the readers but also during the last ten years they turn out to be attractive  being downloaded online. Google Analytics tool provides information about the visiting journal's web-site giving the idea regarding the geographic spreading of the visitors that are potential readers of the journal.

Another topic of interest is connected with citing of CMP papers. Since it is not worth repeating the number that can be obtained from official scientometric services, here we concentrate on the correlation between the downloads and citations of publications. The pairs of indicators (the number of downloads and the number of citations) were found for 320 CMP papers. Plotting these numbers, one can see that besides the bulk of the papers, a dozen of leaders can be listed. It is interesting that these papers can be divided into two groups, i.e., highly cited but less downloaded and highly downloaded but less cited. It is an open question for the future research to understand which factors exactly are responsible for such a segregation: whether these factors concern an interesting description, the topic of the present publication or maybe an additional promotion in social media.

Concluding, it can be definitely stated that CMP journal already now occupies a good position which it definitely deserves. Surely, it also has very good prospects to develop further by strengthening its international collaboration, paying attention to its proper promotion and continuing to attract qualified scholars in order to finally get its completed giant cluster of experts. Undoubtedly, all the above is only natural for the 25-years young journal on the way to its true success. Good luck to you and many happy returns of the jubilee!

\section*{Acknowledgements}
Author thanks the National Academy of Sciences of Ukraine (grant No.~0118U003620), the Ministry of Education and Science of Ukraine (project ``Systematization of high school journals based on scientometric studies'') and the Publishing house ``Akademperiodyka'' (project ``Formation and developing the electronic resources presenting the publishing products of the National Academy of Sciences of Ukraine''). Many thanks also to Yurij Holovatch and Ihor Mryglod for their kind assistance and discussions.

\ukrainianpart
\title{Наукометричний аналіз журналу Condensed Matter Physics}
\author{О. Мриглод}

\addresses{\addr{a1}Інститут фізики конденсованих систем НАН України, вул. Свєнціцького,  1, 79011  Львів,  Україна
\addr{a2}Спiвпраця \textbf{L}$^4$ i Коледж докторантiв зi статистичної фiзики складних систем,
Ляйпцiг–Лотарингiя–Львiв–Ковентрi, Європа%
}

\makeukrtitle

\begin{abstract}
Стаття присвячена 25-му ювілею журналу \emph{Condensed Matter Physics} (CMP).  Тут наведені результати багатоаспектного аналізу різного типу даних, що мають стосунок до журналу. Взаємозв'язки співавторства в рамках CMP досліджуються шляхом аналізу мережі спрівпраці. Обговорюються її кумулятивні статичні та динамічні властивості, а також --- структура. Використання авторських даних про місце праці дає змогу оцінити міжнародний внесок у журнал. Розглядається мережа країн, пов'язаних зв'язками співавторства у CMP. Інший вид мереж використовується для аналізу тематичного спектру: у цьому випадку лінком поєднується пара номерів PACS, якщо вони фігурують в одній статті. Аналізується структура найбільш вагомих міжтематичних зв'язків. Нарешті, статистика завантажень та відповідні записи про цитованість статей застосовуються для оцінки впливу журналу.

\keywords складні системи, складні мережі, наукометрія, оцінювання впливу, оцінювання журналу
\end{abstract}
\end{document}